\documentclass{aa63273-26} 

\usepackage{natbib}
\usepackage{hyperref}
\usepackage[T1]{fontenc}
\usepackage{ae,aecompl}
\usepackage{rotating}
\usepackage{color}
\usepackage{graphicx}
\usepackage{appendix}
\usepackage{longtable}
\usepackage{txfonts}
\usepackage{orcidlink}
\usepackage{float}

\begin{document}

\authorrunning{S. Allak}

\title{NGC~1300: A rich reservoir of ultraluminous X-ray sources}
\titlerunning{A rich reservoir of ULXs in NGC 1300}

\author{ Sinan Allak\orcidlink{0000-0001-7093-1079}\inst{1}, Faruk Soydugan\orcidlink{0000-0002-5141-7645}\inst{2}, Andrea Santangelo\orcidlink{0000-0003-4187-9560}\inst{1}, Wei Yu\orcidlink{0000-0002-3229-2453}\inst{1}, Santina Piraino\orcidlink{0000-0003-0122-6899}\inst{1} \and Daniel Kulinsky\inst{1} }

\institute{ Institut für Astronomie und Astrophysik, Sand 1, 72076 Tübingen, Germany\\ \email{sinan.allak@uni-tuebingen.de} \and Department of Physics, Çanakkale Onsekiz Mart University, 17100 Çanakkale, Türkiye }

\abstract{We present the first systematic multi-epoch study of the X-ray source population of the barred spiral galaxy NGC~1300 using 12 archival \textit{Chandra} observations spanning more than 15 years, complemented by archival \textit{Swift}/XRT monitoring. We identify 42 X-ray sources within the optical extent of the galaxy, including six ultraluminous X-ray sources (ULXs) and 13 ULX candidates. The latter are classified as candidates because their measured luminosities remain below $10^{39}~{\rm erg~s^{-1}}$, while their upper uncertainty bounds reach or exceed this threshold in at least one epoch. We investigate the luminosities, long-term variability, broad-band spectral properties, and epoch-resolved X-ray luminosity functions (XLFs) of the population. Ten sources satisfy our adopted transient criterion, while seven have a variability significance of $S_{\rm var}\geq3$, with four sources satisfying both variability criteria. X32 is a highly variable ULX with a peak X-ray luminosity of $1.5\times10^{40}~{\rm erg~s^{-1}}$, which exceeds the upper limit from the non-detection by a factor greater than 715. The hardness--intensity diagrams reveal diverse spectral behavior among the luminous sources, with no common hardness--luminosity evolution. In particular, X32 remains predominantly hard despite its large luminosity variations, and its X-ray colors are consistent with the empirical high-mass X-ray binary (HMXB) region of the color--color diagram. The cumulative XLFs above $10^{38}~{\rm erg~s^{-1}}$ have slopes of $\beta\simeq0.6$--$1.0$, while evidence for a broken power law is limited to a small number of epochs. The number of off-nuclear sources reaching $10^{39}~{\rm erg~s^{-1}}$ varies from zero to four between epochs, demonstrating that the bright end of the XLF and the observed ULX census are strongly shaped by long-term source variability. Our results show that the observed ULX census of NGC~1300 is strongly epoch dependent and that multi-epoch monitoring is essential for establishing representative ULX populations in nearby galaxies.}

\keywords{X-rays: binaries -- galaxies: individual: NGC 1300 -- X-rays: galaxies -- accretion, accretion disks}

 \maketitle

\section{Introduction}

NGC~1300 is a nearby, strongly barred spiral galaxy of morphological type SB(s)bc \citep{1981rsac.book.....S}, characterized by a prominent stellar bar connected to two well-defined spiral arms. We adopt a distance of 20.7~Mpc, following \citet{2020MNRAS.493.5045M}, corresponding to a physical scale of $100~{\rm pc~arcsec^{-1}}$. This distance is based on a systemic velocity of $1511~{\rm km~s^{-1}}$, corrected for the Virgo Cluster, Great Attractor, and Shapley concentration, and $H_0=73~{\rm km~s^{-1}~Mpc^{-1}}$. The galaxy has an inclination of $50.2^\circ$ \citep{1989ApJ...337..191E}. NGC~1300 is actively forming stars, with a global star-formation rate (SFR) of $1.4~M_\odot\,{\rm yr^{-1}}$ when rescaled to the distance adopted here, and has a stellar mass of $\log(M_\star/M_\odot)\simeq10.6$ \citep{2021ApJS..257...43L}. Its prominent bar, spiral arms, and bar-end regions contain markedly different molecular-gas and star-forming environments \citep{2020MNRAS.493.5045M,2020MNRAS.495.3840M}. Spatially resolved stellar-population analysis further indicates sub-solar metallicity across much of NGC~1300, with mean $[{\rm M/H}]$ values of $-0.22$ dex in the disk and $-0.18$ dex in the nuclear ring, compared with approximately solar metallicity in the nucleus \citep{2020A&A...644A.116R}. These spatially distinct environments make NGC~1300 a useful system for investigating whether variations in local star-forming conditions are reflected in the properties and distribution of its compact X-ray source population.

The molecular gas and star-forming environments of NGC~1300 have been studied extensively through CO observations. Single-dish $^{12}{\rm CO}(1$--$0)$ observations revealed substantial molecular gas in both the bar and spiral arms, despite the lack of prominent H\,{\sc ii} regions along much of the bar \citep{2018PASJ...70...37M}. High-resolution ALMA observations subsequently identified 233 giant molecular clouds (GMCs) in the bar, spiral arms, and bar-end regions, with no strong systematic differences in their sizes, velocity dispersions, molecular masses, or virial parameters \citep{2020MNRAS.493.5045M}. Despite these similarities in GMC properties, star formation is strongly suppressed in the bar. The molecular-gas star-formation efficiency is $0.06$--$0.16~{\rm Gyr^{-1}}$ in the bar, compared with $0.23$--$0.96~{\rm Gyr^{-1}}$ in the spiral arms and bar-end regions, while the diffuse molecular-gas fraction reaches $0.74$--$0.91$ in the bar \citep{2020MNRAS.495.3840M}. Hydrodynamical simulations tailored to NGC~1300 suggest that the elongated gas motions induced by the strong bar increase cloud--cloud collision velocities and may contribute to the suppression of massive star formation \citep{2020MNRAS.494.2131F}. The resulting contrast between the relatively quiescent bar and the more actively star-forming spiral arms and bar-end regions is particularly relevant to its X-ray source population, because high-mass X-ray binaries (HMXBs) are generally associated with recent star formation, whereas low-mass X-ray binaries (LMXBs) more closely trace the older stellar population \citep{2006ARA&A..44..323F}.

X-ray binaries (XRBs) constitute a substantial fraction of the resolved X-ray source populations of nearby galaxies. The numbers and integrated X-ray luminosities of HMXBs scale with the recent SFR, whereas LMXBs predominantly trace stellar mass \citep{2004MNRAS.349..146G,2006ARA&A..44..323F, 2012MNRAS.419.2095M,2019ApJS..243....3L}. The X-ray luminosity function (XLF) is therefore commonly used to characterize resolved XRB populations and their dependence on the stellar content and star-formation activity of the host galaxy. In star-forming galaxies, HMXBs dominate the luminous end of the XLF, which extends into the ultraluminous X-ray source (ULX) regime \citep{2003MNRAS.339..793G,2012MNRAS.419.2095M}. ULXs are non-nuclear X-ray sources that reach $L_{\rm X}\geq10^{39}~{\rm erg~s^{-1}}$ and include neutron stars and stellar-mass black holes accreting at high or super-Eddington rates \citep{2014Natur.514..202B,2017ARA&A..55..303K, 2021AstBu..76....6F}. Their observed luminosities can vary by several orders of magnitude, and individual sources may reach across the conventional ULX threshold only during their brightest states. Repeated observations have revealed transient ULXs that were either undetected or observed at substantially lower luminosities in earlier epochs, including systems in NGC~5907, NGC~7090, NGC~55, and NGC~4254 \citep{2018MNRAS.477L..90P,2020ApJ...891..153E, 2021MNRAS.501.1002W,2022MNRAS.515.4669R, 2023MNRAS.526.5765A}. ULX censuses based on single-epoch observations may not capture sources that attain ultraluminous luminosities only intermittently. Multi-epoch X-ray observations are required to sample the luminosity range of individual sources, identify transient systems, and determine how strongly variable sources affect both the observed ULX census and the bright end of the XLF.

The X-ray source population of NGC~1300 has received comparatively little attention. Using a single $\sim30$~ks \textit{Chandra} ACIS-S observation obtained as part of a survey of galaxies hosting dynamically measured supermassive black holes, \citet{2012ApJ...749..129G} reported 17 X-ray point sources projected within the galaxy, including the nuclear source and three ULX candidates. However, the X-ray source population of NGC~1300 has not previously been investigated systematically over multiple epochs. In this work, we present the first systematic multi-epoch study of the X-ray source population of NGC~1300 using 12 archival \textit{Chandra} ACIS-S observations spanning more than 15 years. Source detection is performed independently at each epoch, allowing us to construct a multi-epoch source catalog and follow the luminosity evolution of individual sources. We investigate long-term variability and transient behavior and characterize broad-band spectral changes using hardness ratios and X-ray color diagnostics. We also construct the observed XLF at each \textit{Chandra} epoch to examine changes in the luminosity distribution with time. Archival \textit{Swift}/XRT observations provide additional long-term coverage. Particular attention is given to the ULX population and to X32, the most luminous source in our sample, for which the substantial luminosity variability and multi-epoch coverage allow its long-term luminosity and spectral behavior to be examined in greater detail. The multi-epoch data allow us to assess how source variability affects the observed ULX population and the bright end of the XLF in NGC~1300.

\section{Observations, data reduction, and analysis} \label{DRA}

\subsection{Observations}

This work is based on 12 archival observations of NGC\,1300 obtained with the \textit{Chandra X-ray Observatory} using the Advanced CCD Imaging Spectrometer (ACIS-S). The observations span from October 2009 to August 2025, providing a long temporal baseline for investigating the X-ray source population and its long-term variability. The first observation (ObsID 11775) was obtained in 2009, while the remaining 11 observations were carried out between December 2024 and August 2025. The individual exposure times range from 8.82 to 53.35~ks. The observation IDs, exposure times, and start dates are listed in Table~\ref{tab:obs}.

\subsection{Chandra data reduction and analysis}

The \textit{Chandra} ACIS-S observations were processed using the \textit{Chandra} Interactive Analysis of Observations (CIAO; \citealt{2006SPIE.6270E..1VF}) together with CALDB v4.11, and reprocessed level-2 event files were created with the standard \textit{chandra\_repro} task. Source detection was performed independently for each of the 12 observations in the 0.3--10~keV band and was restricted to the optical extent of NGC~1300 defined by its $D_{25}$ ellipse, centered at $\alpha_{\rm J2000}=03^{\rm h}19^{\rm m}41.1525^{\rm s}$ and $\delta_{\rm J2000}=-19^{\circ}24^{\prime}41.734^{\prime\prime}$, with semi-axes of $186^{\prime\prime}$ and $123^{\prime\prime}$ and a position angle of $106.0^{\circ}$ (see Fig.\ref{fig:rgbchandra}). Exposure-corrected images and corresponding PSF maps were generated with \textit{fluximage} using the native ACIS pixel size (\textit{binsize=1}), a representative energy of 2.3~keV, and a 90\% encircled-count fraction (\textit{psfecf=0.9}). A blind source search was then performed with \textit{wavdetect}, without supplying prior source positions, using wavelet scales of 1, 1.414, 2, 2.828, 4, 5.657, 8, 11.314, and 16 pixels and a detection threshold of \textit{sigthresh=$10^{-6}$}; the corresponding exposure and PSF maps were supplied to account for spatial variations in sensitivity and source extent across the detector. The same detection procedure was applied uniformly to all epochs, while sources were not required to be detected in every observation. This approach allowed detections and non-detections to be tracked across epochs and enabled us to investigate the long-term behavior of the X-ray source population within the $D_{25}$ region. Detections from the individual observations were cross-matched by position using a matching radius of $1\arcsec$ to construct a master source catalog. Sources detected in multiple observations were assigned a single source identifier (X1--X42), while detections without a positional counterpart in the other epochs were retained as individual catalog sources.

Source spectra and light curves were extracted from the reprocessed event files using the CIAO tasks \textit{specextract} and \textit{dmextract}, respectively. Source photons were extracted from circular apertures with radii of up to $4^{\prime\prime}$. The local background was generally estimated from concentric annuli with inner and outer radii of $6^{\prime\prime}$ and $12^{\prime\prime}$, respectively. For X14/X22, X8/X10, and X16/X33, we used smaller source apertures and adjusted the background annulus radii to reduce contamination from nearby sources. The background regions were visually inspected and adjusted when necessary to minimize contamination from neighboring point sources and regions of enhanced diffuse X-ray emission. The \textit{specextract} task was used to generate the source and background spectra together with the corresponding redistribution matrix files (RMFs) and ancillary response files (ARFs), thereby accounting for the instrumental response at the source position. Because the available photon statistics vary substantially among the detected sources and between individual epochs, the spectra were grouped to contain 5--10 counts per spectral bin, depending on the number of source counts available. Spectral fitting was performed using the Cash statistic (C-stat), which is better suited to Poisson-dominated X-ray spectra than $\chi^{2}$ minimization in the low-count regime. The extracted light curves were background-subtracted using the corresponding local background regions and were used, together with the multi-epoch flux measurements, to examine temporal variability on both individual-observation and long-term timescales.

\subsection{X-ray fluxes and luminosities}

Following source detection, X-ray fluxes were estimated independently for each source in each of the 12 \textit{Chandra} observations using the CIAO task \textit{srcflux}. Since a substantial fraction of the detected sources have limited photon statistics, conventional spectral extraction and fitting do not provide sufficiently reliable constraints for the full source sample. We therefore adopted \textit{srcflux}, which provides a uniform approach for estimating source fluxes and confidence intervals in the low-count regime while accounting for the local background and instrumental response. Fluxes were calculated in three energy bands: 0.3--10~keV (total), 0.3--2~keV (soft), and 2--10~keV (hard). For the count-rate-to-flux conversion, we adopted an absorbed power-law model with a photon index fixed at $\Gamma=1.7$, as in previous ULX surveys of nearby galaxies \citep{2006ApJ...642..171L,2020MNRAS.491.1260S}, and a Galactic hydrogen column density fixed at $N_{\rm H}=2\times10^{20}~{\rm cm^{-2}}$. We used the same photon index for all sources and epochs because many detections have too few counts to constrain it independently. Aperture corrections were applied using the \textit{arfcorr} method, and uncertainties were calculated at the 68.27\% confidence level. For observations in which a source was not significantly detected, the corresponding \textit{srcflux} constraints were retained as upper limits where available, allowing both detections and non-detections to be incorporated into the long-term analysis. The resulting absorption-corrected fluxes were converted to intrinsic X-ray luminosities using $L_{\rm X}=4\pi d^{2}F_{\rm X}$, adopting a distance of 20.7~Mpc to NGC~1300. 

\subsection{X-ray photometry and spectral hardness}

Using the soft- and hard-band flux measurements described above, the broad-band spectral hardness of each source was characterized as ${\rm HR}=(H-S)/(H+S)$, where $S$ and $H$ correspond to the 0.3--2 and 2--10~keV fluxes, respectively. The asymmetric uncertainties on HR were propagated from the corresponding lower and upper uncertainties returned by \textit{srcflux}. Hardness ratios were calculated only for epochs with meaningful measurements in both energy bands. To examine long-term spectral changes and their possible dependence on source luminosity, hardness--intensity diagrams were constructed by comparing HR with the corresponding 0.3--10~keV luminosity. When either energy band has few counts, large HR uncertainties can obscure changes in hardness between epochs. To exclude poorly constrained measurements while retaining sufficient temporal coverage, we required both the lower and upper HR uncertainties to be $\leq 0.40$. We included only sources with at least four epochs that met this criterion in the hardness--intensity diagrams.

For X32, which provided sufficiently well-constrained measurements for a more detailed color analysis, we additionally followed the X-ray color--color scheme of \citet{2003ApJ...595..719P}. We repeated the \textit{srcflux} measurements in the $S=0.3$--1.0~keV, $M=1.0$--2.0~keV, and $H=2.0$--8.0~keV bands, adopting the same spectral model, confidence level, and aperture correction described above. The soft and hard colors were defined as ${\rm SC}=(M-S)/(S+M+H)$ and ${\rm HC}=(H-M)/(S+M+H)$, respectively. Only epochs with constrained \textit{srcflux} measurements in all three bands were used in the color--color diagram, and the resulting colors were compared with the empirical source regions defined by \citet{2003ApJ...595..719P}. To further investigate the nature of the compact object in X32, we calculated the softness ($F_{2-4}/F_{4-6}$) and hardness ($F_{6-30}/F_{4-6}$) ratios following \citet{2017ApJ...836..113P}. The 6--30~keV fluxes were extrapolated from the best-fitting spectral models using XSPEC.

\subsection{Timing and long-term variability analysis}

A systematic timing analysis was performed for the detected X-ray sources whenever the available photon statistics were sufficient. For sources with more than 100 source photons in an individual \textit{Chandra} observation, we searched for coherent periodic signals using the $Z_n^2$ test. The event arrival times were corrected to the Solar System barycenter prior to the timing analysis. Background-subtracted light curves were also generated with several temporal bin sizes to examine short-term intensity variations and to assess whether candidate timing features were associated with changes in the source count rate. In addition to the intra-observation timing analysis, we investigated long-term variability using the 0.3--10~keV luminosities measured in the 12 \textit{Chandra} observations. The resulting multi-epoch \textit{Chandra} light curves for six selected sources (X17, X19, X21, X23, X28, and X29) are presented in Fig.~\ref{fig:longterm_lc}. These sources were selected because their luminosities are comparatively well constrained across the observed epochs, allowing their long-term variability to be clearly traced. To quantify long-term variability, we calculated the variability significance as \begin{equation} S_{\rm var} = \frac{L_{\rm X,max}-L_{\rm X,min}} {\sqrt{\sigma_{\rm max}^{2}+\sigma_{\rm min}^{2}}}, \end{equation} where $L_{\rm X,max}$ and $L_{\rm X,min}$ are the maximum and minimum measured luminosities, respectively, and $\sigma_{\rm max}$ and $\sigma_{\rm min}$ are their corresponding uncertainties. Sources with $S_{\rm var}\geq3$ were considered significantly variable.  We classified a source as transient if it was undetected in at least one epoch and showed a luminosity ratio \begin{equation} R=\frac{L_{\rm X,max}}{L_{\rm X,UL}}\geq10, \end{equation} where $L_{\rm X,UL}$ is the minimum luminosity upper limit among the non-detections.

NGC~1300 has also been observed on multiple occasions with the \textit{Neil Gehrels Swift Observatory} X-Ray Telescope (\textit{Swift}/XRT; Table~\ref{tab:obs}), providing an additional temporal baseline beyond the \textit{Chandra} observations. We searched the \textit{Swift}/XRT data at the positions of the \textit{Chandra} sources that could be spatially distinguished at the angular resolution of XRT (X17, X19, X23, X28, and X32; Fig.~\ref{fig:xrt}). The XRT products were generated using the online \textit{Swift}/XRT data product generator \citep{2007A&A...469..379E,2009MNRAS.397.1177E}. For each observation, the resulting XRT count rates were converted to unabsorbed 0.3--10~keV fluxes using WebPIMMS, adopting the same absorbed power-law model used for the \textit{Chandra} \textit{srcflux} analysis. The unabsorbed fluxes were subsequently converted to X-ray luminosities assuming a distance of 20.7~Mpc. For epochs without a significant XRT detection, 3$\sigma$ upper limits were calculated and converted to luminosity limits using the same procedure. Among the sources that could be distinguished with XRT, X32 was detected in several \textit{Swift}/XRT observations, while the remaining observations yielded upper limits. Because of the broader XRT PSF, the \textit{Swift}/XRT measurements were interpreted with caution where nearby \textit{Chandra} sources could contribute to the XRT extraction region.

\subsection{X-ray luminosity functions and background AGN contamination}
\label{sec:xlf_method}

We constructed cumulative X-ray luminosity functions (XLFs) separately for each of the 12 \textit{Chandra} observations in order to examine epoch-to-epoch changes in the luminosity distribution of the source population. The nuclear source X12 was excluded from the XLF analysis. For each observation, only sources with a measured 0.3--10~keV luminosity were included, while non-detections and their corresponding upper limits were excluded from the cumulative source counts. The cumulative XLF was defined as the number of detected sources with luminosities greater than a given luminosity, $N(>L_{\rm X})$. To minimize the influence of the poorly sampled faint end, the quantitative fits were restricted to sources with $L_{\rm X}\geq10^{38}~{\rm erg~s^{-1}}$.

We first described each XLF with a single power law, $N(>L_{\rm X})\propto L_{\rm X}^{-\beta}$, where $\beta$ is the cumulative XLF slope. The slope was determined from the unbinned luminosity measurements using the maximum-likelihood estimator for a power-law distribution above the adopted lower luminosity threshold. For a differential distribution $p(L_{\rm X})\propto L_{\rm X}^{-\alpha}$, the maximum-likelihood estimate is \begin{equation} \alpha = 1+\frac{N} {\displaystyle\sum_{i=1}^{N} \ln(L_{{\rm X},i}/L_{\rm min})}, \end{equation} where $L_{\rm min}=10^{38}~{\rm erg~s^{-1}}$ and $\beta=\alpha-1$. We also fitted a continuous broken power-law model ($N(>L_{\rm X})\propto L_{\rm X}^{-\beta_1}$ for $L_{\rm X}<L_{\rm b}$ and $N(>L_{\rm X})\propto L_{\rm X}^{-\beta_2}$ for $L_{\rm X}\geq L_{\rm b}$.), allowing the slopes below and above the break and the break luminosity, $L_{\rm b}$, to vary. The single- and broken-power-law descriptions were compared using the Bayesian information criterion (BIC). We defined \begin{equation} \Delta{\rm BIC} = {\rm BIC}_{\rm single}-{\rm BIC}_{\rm broken}, \end{equation} such that positive values indicate a preference for the broken power-law model. No completeness correction was applied to the epoch-resolved XLFs; consequently, the fitted distributions represent the observed source population above the adopted luminosity threshold. Moreover, for comparison with the expected HMXB population, we adopted the SFR-normalized HMXB XLF of \citet{2012MNRAS.419.2095M}. The SFR of NGC~1300 was rescaled to the adopted distance of 20.7~Mpc, giving ${\rm SFR}=1.4~M_\odot\,{\rm yr}^{-1}$. The 0.5--8~keV luminosity scale of the HMXB XLF was converted to our 0.3--10~keV band assuming a power-law spectrum with $\Gamma=1.7$, consistent with the spectral model adopted for the source flux estimates. The intrinsic 0.34-dex scatter in the HMXB XLF normalization was included when comparing the expected and observed numbers of sources above $10^{39}~{\rm erg~s^{-1}}$.

We independently estimated the expected contamination by unrelated background active galactic nuclei (AGNs). For this purpose, the individual \textit{Chandra} observations were reprojected and combined in the 2--10~keV band. Exposure and 90\% encircled-count-fraction PSF maps were generated for the merged data, and a background map was constructed from the merged hard-band counts image. These products were used to calculate a limiting-sensitivity map, $S_{\rm lim}(x,y)$, which accounts for the spatial variations in exposure, background, and PSF across the field. The calculation was restricted to the $D_{25}$ ellipse used for the source selection, with an area of approximately $19.96~{\rm arcmin^2}$ ($5.55\times10^{-3}~{\rm deg^2}$). The limiting photon flux at each position was converted to a 2--10~keV energy flux assuming a power-law spectrum with $\Gamma=1.7$, consistent with that adopted for the source flux estimates.  The expected surface density of background sources was calculated using the 2--10~keV $\log N$--$\log S$ relation of \citet{2009A&A...497..635C}. They parameterized the differential source counts as a broken power law, $n(S)=dN/dS=A S^{-\alpha_1}$ for $S>S_{\rm b}$ and $n(S)=dN/dS=B S^{-\alpha_2}$ for $S\leq S_{\rm b}$, with $A=B S_{\rm b}^{\alpha_1-\alpha_2}$, where $S$ is expressed in units of $10^{-14}~{\rm erg~cm^{-2}~s^{-1}}$. For the 2--10~keV band, \citet{2009A&A...497..635C} obtained $\alpha_1=2.46\pm0.08$, $\alpha_2=1.55\pm0.18$, and $S_{\rm b}=(1.05\pm0.16)\times10^{-14}~ {\rm erg~cm^{-2}~s^{-1}}$, with a normalization of $A=413$. At each position within the $D_{25}$ ellipse, the cumulative surface density of background sources brighter than the local limiting flux was obtained by integrating this differential relation above $S_{\rm lim}(x,y)$. The total expected number of unrelated background sources was then calculated by integrating this surface density over the $D_{25}$ area, $N_{\rm AGN}=\int_{D_{25}}N[>S_{\rm lim}(x,y)]\,d\Omega$. This procedure yields an expectation of approximately 16 background sources within the $D_{25}$ region. The AGN estimate was used to quantify the statistical level of background contamination and was not subtracted source-by-source from the observed XLFs.

\section{Results} \label{R}

\begin{figure}
\resizebox{\hsize}{!}{\includegraphics{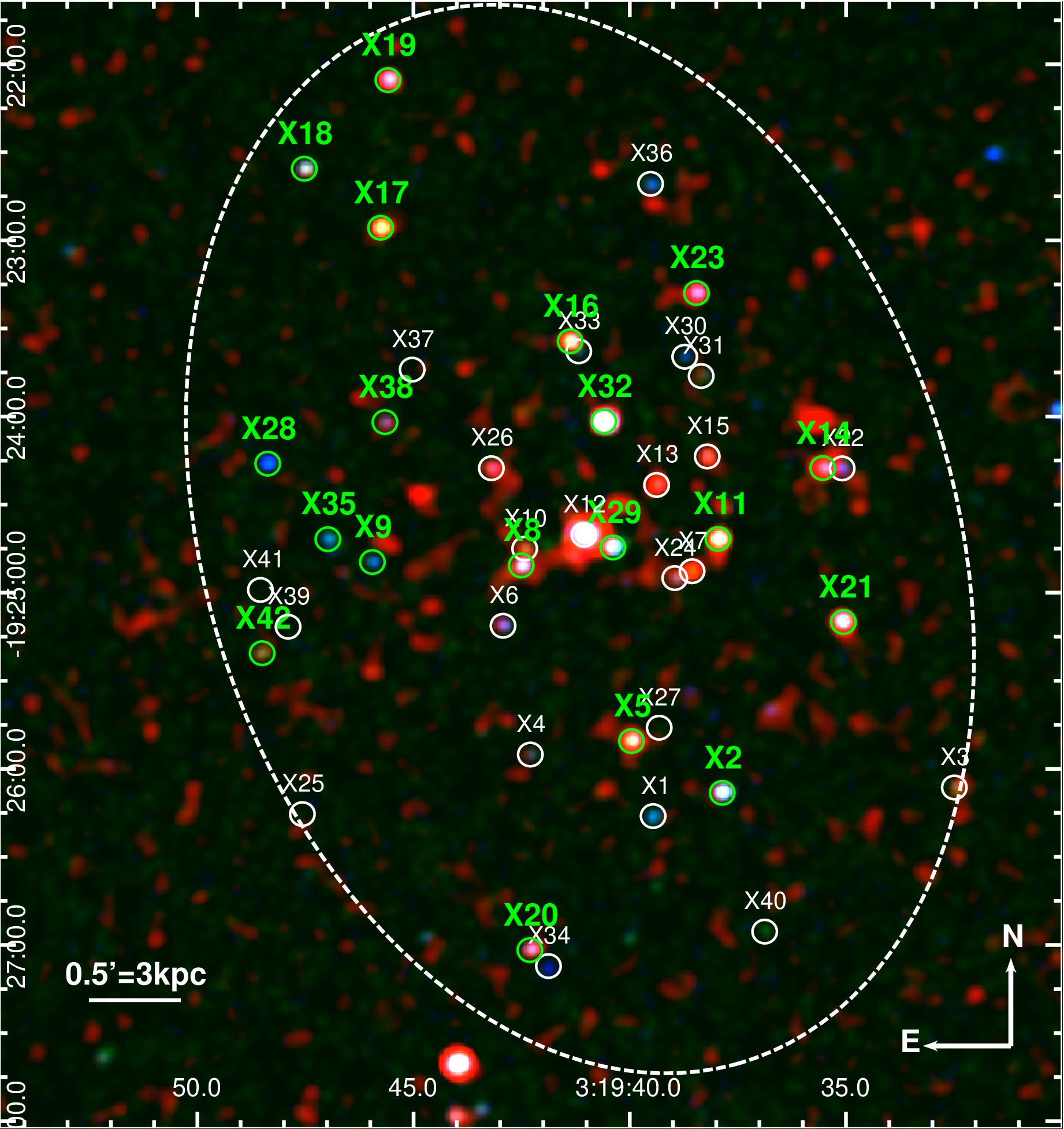}}
\caption{Three-color \textit{Chandra} X-ray image of NGC~1300 constructed from the stacked observations. Red, green and blue correspond to the 0.3--1, 1--2 and 2--10~keV energy bands, respectively. The image is centered on source X12 at R.A.$=03^{\rm h}19^{\rm m}41.06^{\rm s}$ and Dec.$=-19^\circ24'40.31''$ (J2000). X-ray sources detected within the adopted optical extent of NGC~1300 are marked and labeled X1--X42. Sources classified as secure ULXs and ULX candidates are highlighted in green, while the remaining detected sources are shown in white. The large dashed ellipse denotes the adopted $D_{25}$ region of the galaxy. North is up and east is to the left, as indicated by the orientation marker in the lower-right corner. The scale bar corresponds to $0.5'$ ($\simeq3.0$~kpc at a distance of 20.7~Mpc).}
\label{fig:rgbchandra}
\end{figure}

Source detection was performed independently for each of the 12 \textit{Chandra} observations of NGC~1300. We identified a total of 42 X-ray sources within the $D_{25}$ region of the galaxy, including one source (X12) associated with the nucleus of NGC~1300. Figure~\ref{fig:rgbchandra} shows the spatial distribution of the detected sources on the stacked three-color \textit{Chandra} image. The coordinates of all detected X-ray sources are listed in Table~\ref{tab:xray_sources}. We cross-matched the X-ray source positions with the 2MASS, Gaia DR3, SDSS, and radio source catalogs to search for possible multiwavelength counterparts and potential foreground or background contaminants. X22 has a positional counterpart in 2MASS, while X14 is matched with a Gaia DR3 source; no other secure catalog matches were identified. Among the 41 off-nuclear sources, six are classified as ULXs, 13 as ULX candidates, 20 as bright sources, and two as faint sources. Ten sources satisfy our transient criterion, while seven sources have $S_{\rm var}\geq3$; four sources (X7, X13, X29, and X32) satisfy both variability criteria. As shown in Fig.~\ref{fig:jwst_rgb}, most of the X-ray sources are projected onto the disk of NGC~1300, with a substantial fraction located along or close to the prominent spiral-arm structures. The derived X-ray properties of these sources, including long-term variability, spectral hardness, and luminosity distributions, are presented in detail in the following subsections. 

\subsection{X-ray luminosities}

The 0.3--10~keV X-ray luminosities derived for the 12 \textit{Chandra} observations are presented in Table~\ref{tab:lx_allobs}. The measured luminosities span more than three orders of magnitude, from approximately $1\times10^{37}$ to $1.46\times10^{40}~{\rm erg~s^{-1}}$. Most sources occupy the $10^{37}$--$10^{38}~{\rm erg~s^{-1}}$ luminosity range, whereas only a small fraction reach luminosities above $10^{39}~{\rm erg~s^{-1}}$. X32 is the most luminous source in the sample, reaching $L_{\rm X}=1.46\times10^{40}~{\rm erg~s^{-1}}$ in ObsID~28120. X29 and X21 reach maximum luminosities of $2.44\times10^{39}$ and $1.84\times10^{39}~{\rm erg~s^{-1}}$, respectively. X11 and X18 both reach $1.08\times10^{39}~{\rm erg~s^{-1}}$, while X2 reaches $1.05\times10^{39}~{\rm erg~s^{-1}}$. The nuclear source X12 remains above $10^{39}~{\rm erg~s^{-1}}$ in all 12 observations, spanning $L_{\rm X}=(1.85$--$5.72)\times10^{39}~{\rm erg~s^{-1}}$. Based on these luminosities, six off-nuclear sources (X2, X11, X18, X21, X29, and X32) reach or exceed $10^{39}~{\rm erg~s^{-1}}$ in at least one observation and are classified as ULXs. Thirteen additional sources (X5, X8, X9, X14, X16, X17, X19, X20, X23, X28, X35, X38, and X42) remain below $10^{39}~{\rm erg~s^{-1}}$ in their measured luminosities, but their upper uncertainty bounds reach or exceed this threshold in at least one epoch; these are therefore classified as ULX candidates. Because of its nuclear location, X12 is excluded from the census. Accordingly, the \textit{Chandra} observations identify six ULXs and 13 additional ULX candidates within the $D_{25}$ region of NGC~1300. The remaining off-nuclear population consists of 20 bright sources and two faint sources, X25 and X27. In addition, we cross-matched our sources with the Chandra catalog of \cite{2012ApJ...749..129G} to compare our luminosity classifications with those reported in previous work. Their sources 6, 9, and 12 correspond to X7, X24, and X5 in our catalog and were reported as candidate ULXs. However, their absorption-corrected luminosities were obtained from spectral fits in which both $N_{\rm H}$ and $\Gamma$ were free. The resulting steep photon indices ($\Gamma=3.8$--5.9) produced large and poorly constrained absorption corrections. When $N_{\rm H}$ was fixed to the Galactic value, the luminosities decreased to $(2.3$--$7.8)\times10^{38}~{\rm erg~s^{-1}}$, consistent with the luminosity range found in our analysis. The previously reported ULX classifications of these sources are therefore strongly dependent on the adopted spectral model.

\begin{sidewaystable*}
\centering
\fontsize{9}{10}\selectfont
\caption{X-ray luminosities of the NGC~1300 sources in the 12 \textit{Chandra} observations.}
\label{tab:lx_allobs}

\setlength{\tabcolsep}{2.4pt}
\begin{tabular}{lcccccccccccc}
\hline\hline
Source & 11775 & 28537 & 30697 & 30698 & 28539 & 28541 & 30743 & 28538 & 30963 & 28540 & 30966 & 28120 \\
\hline
X1 & $2.58_{-0.66}^{+0.79}$ & $1.37_{-1.13}^{+2.35}$ & $2.65_{-1.61}^{+2.65}$ & $1.22_{-1.11}^{+2.27}$ & $1.50_{-0.78}^{+1.17}$ & $0.71_{-0.60}^{+1.21}$ & $0.99_{-0.78}^{+1.62}$ & $1.13_{-1.03}^{+2.09}$ & $1.14_{-0.83}^{+1.32}$ & $1.27_{-0.85}^{+1.39}$ & $<1.16$ & $2.16_{-0.77}^{+0.95}$ \\
X2 & $6.29_{-1.02}^{+1.16}$ & $2.89_{-1.69}^{+2.74}$ & $2.54_{-1.57}^{+2.60}$ & $6.11_{-2.72}^{+3.87}$ & $4.94_{-1.54}^{+1.94}$ & $6.18_{-2.13}^{+2.75}$ & $7.46_{-2.52}^{+3.25}$ & $9.54_{-3.29}^{+4.25}$ & $10.50_{-2.58}^{+3.08}$ & $6.41_{-2.00}^{+2.51}$ & $9.50_{-2.79}^{+3.51}$ & $8.02_{-1.41}^{+1.60}$ \\
X3 & $1.52_{-0.54}^{+0.67}$ & $<1.94$ & $0.99_{-0.99}^{+2.16}$ & $1.28_{-1.23}^{+2.47}$ & $1.98_{-1.00}^{+1.43}$ & $0.46_{-0.46}^{+1.28}$ & $1.07_{-0.84}^{+1.74}$ & $<1.68$ & $<0.81$ & $0.36_{-0.36}^{+1.09}$ & $1.70_{-1.15}^{+1.86}$ & $1.45_{-0.65}^{+0.84}$ \\
X4 & $1.99_{-0.56}^{+0.69}$ & -- & $<1.59$ & $1.36_{-1.12}^{+2.32}$ & $0.40_{-0.39}^{+0.78}$ & $<0.93$ & -- & $1.15_{-1.10}^{+2.22}$ & $1.36_{-0.86}^{+1.41}$ & $2.22_{-1.14}^{+1.69}$ & $1.94_{-1.22}^{+1.97}$ & $0.25_{-0.25}^{+0.50}$ \\
X5 & $5.29_{-0.94}^{+1.07}$ & $1.36_{-1.08}^{+2.23}$ & $6.76_{-2.68}^{+3.67}$ & $1.57_{-1.24}^{+2.57}$ & $0.92_{-0.63}^{+1.02}$ & $1.54_{-0.96}^{+1.58}$ & $0.91_{-0.79}^{+1.61}$ & $3.93_{-2.02}^{+2.99}$ & $3.36_{-1.36}^{+1.85}$ & $4.16_{-1.55}^{+2.05}$ & $4.39_{-1.90}^{+2.58}$ & $5.51_{-1.23}^{+1.44}$ \\
X6 & $2.45_{-0.62}^{+0.74}$ & $1.26_{-1.09}^{+2.23}$ & $1.18_{-0.98}^{+2.02}$ & $5.73_{-2.48}^{+3.49}$ & $2.07_{-0.91}^{+1.30}$ & $0.67_{-0.58}^{+1.19}$ & $2.05_{-1.24}^{+2.05}$ & $4.06_{-1.99}^{+2.97}$ & $2.04_{-1.06}^{+1.59}$ & $<0.86$ & $<1.15$ & $2.11_{-0.70}^{+0.90}$ \\
X7 & $4.44_{-0.83}^{+0.96}$ & -- & $<1.60$ & -- & $<0.65$ & $<0.97$ & $0.95_{-0.79}^{+1.63}$ & -- & $<0.78$ & $0.53_{-0.51}^{+1.01}$ & $<1.09$ & $<0.30$ \\
X8 & $3.60_{-0.74}^{+0.87}$ & -- & $2.45_{-1.56}^{+2.55}$ & $9.04_{-3.29}^{+4.39}$ & $4.80_{-1.47}^{+1.85}$ & $4.76_{-1.73}^{+2.31}$ & $6.16_{-2.34}^{+3.08}$ & $6.99_{-2.77}^{+3.80}$ & $3.35_{-1.41}^{+1.91}$ & $4.99_{-1.75}^{+2.26}$ & $6.68_{-2.30}^{+2.98}$ & $4.46_{-1.05}^{+1.23}$ \\
X9 & $2.38_{-0.62}^{+0.74}$ & $5.88_{-2.64}^{+3.76}$ & $6.70_{-2.73}^{+3.70}$ & $3.00_{-1.82}^{+3.00}$ & $2.04_{-0.97}^{+1.35}$ & $5.61_{-1.93}^{+2.50}$ & $<1.28$ & $<1.85$ & $<0.91$ & $<0.94$ & $<1.22$ & $<0.31$ \\
X10 & $1.35_{-0.44}^{+0.55}$ & $<1.77$ & $1.27_{-0.96}^{+2.00}$ & $<1.80$ & $0.39_{-0.38}^{+0.76}$ & $0.72_{-0.55}^{+1.14}$ & $0.77_{-0.77}^{+1.54}$ & $1.23_{-1.02}^{+2.10}$ & $1.23_{-0.81}^{+1.30}$ & $4.25_{-1.55}^{+2.03}$ & $<1.16$ & $0.42_{-0.29}^{+0.49}$ \\
X11 & $9.27_{-1.35}^{+1.36}$ & $6.12_{-2.69}^{+3.84}$ & $2.79_{-1.65}^{+2.69}$ & $8.00_{-3.13}^{+4.31}$ & $6.85_{-1.85}^{+2.26}$ & $9.54_{-2.63}^{+3.25}$ & $6.62_{-2.51}^{+3.31}$ & $10.80_{-3.47}^{+4.41}$ & $8.26_{-2.20}^{+2.64}$ & $2.02_{-1.03}^{+1.55}$ & $4.66_{-1.89}^{+2.57}$ & $5.98_{-1.19}^{+1.38}$ \\
X12 & $20.10_{-1.93}^{+1.95}$ & $36.40_{-7.08}^{+8.06}$ & $40.70_{-7.14}^{+8.13}$ & $57.20_{-8.89}^{+10.00}$ & $44.60_{-4.97}^{+5.02}$ & $47.50_{-6.28}^{+6.35}$ & $49.60_{-7.47}^{+7.54}$ & $37.00_{-6.79}^{+7.73}$ & $23.30_{-3.92}^{+4.37}$ & $18.50_{-3.43}^{+3.91}$ & $30.60_{-5.13}^{+5.84}$ & $35.50_{-3.08}^{+3.11}$ \\
X13 & $4.45_{-0.85}^{+0.97}$ & -- & $<1.70$ & $<1.85$ & $<0.66$ & $<1.01$ & $<1.30$ & $<1.58$ & $0.49_{-0.49}^{+0.98}$ & $<0.82$ & -- & $<0.29$ \\
X14 & $2.85_{-0.70}^{+0.84}$ & $3.39_{-2.13}^{+3.44}$ & $6.20_{-2.73}^{+3.89}$ & $3.02_{-2.03}^{+3.30}$ & $0.83_{-0.69}^{+1.10}$ & $1.28_{-1.03}^{+1.63}$ & $3.37_{-1.73}^{+2.56}$ & $2.55_{-1.54}^{+2.55}$ & $2.02_{-1.03}^{+1.55}$ & $1.98_{-1.06}^{+1.60}$ & $0.88_{-0.73}^{+1.51}$ & $1.44_{-0.58}^{+0.76}$ \\
X15 & $2.06_{-0.57}^{+0.68}$ & $3.06_{-1.90}^{+3.14}$ & $2.84_{-1.68}^{+2.74}$ & $<1.91$ & $0.94_{-0.67}^{+1.10}$ & $1.61_{-1.00}^{+1.65}$ & $<1.35$ & $1.17_{-1.01}^{+2.07}$ & $<0.81$ & $0.58_{-0.53}^{+1.09}$ & $0.87_{-0.67}^{+1.37}$ & $1.19_{-0.50}^{+0.68}$ \\
X16 & $5.03_{-0.90}^{+1.03}$ & $1.31_{-1.15}^{+2.34}$ & $1.24_{-1.03}^{+2.13}$ & $4.45_{-2.37}^{+3.49}$ & $5.04_{-1.63}^{+2.04}$ & $7.12_{-2.33}^{+2.93}$ & $1.90_{-1.41}^{+2.28}$ & $3.86_{-1.89}^{+2.82}$ & $1.22_{-0.80}^{+1.30}$ & $1.29_{-0.82}^{+1.34}$ & $5.72_{-2.08}^{+2.78}$ & $4.07_{-0.97}^{+1.14}$ \\
X17 & $7.02_{-1.34}^{+1.53}$ & $7.51_{-3.05}^{+4.14}$ & $5.26_{-2.40}^{+3.42}$ & $4.40_{-2.34}^{+3.45}$ & $7.70_{-1.96}^{+2.35}$ & $7.02_{-2.22}^{+2.79}$ & $1.90_{-1.28}^{+2.08}$ & $1.23_{-1.07}^{+2.18}$ & $1.28_{-0.84}^{+1.36}$ & $2.97_{-1.31}^{+1.86}$ & $3.80_{-1.71}^{+2.39}$ & $2.97_{-0.80}^{+0.98}$ \\
X18 & $4.94_{-1.05}^{+1.23}$ & $10.80_{-3.67}^{+4.79}$ & $3.98_{-2.07}^{+3.11}$ & $1.32_{-1.16}^{+2.37}$ & $2.47_{-1.14}^{+1.55}$ & $4.00_{-1.73}^{+2.35}$ & $2.37_{-1.66}^{+2.39}$ & $2.53_{-1.75}^{+2.84}$ & $0.40_{-0.40}^{+1.14}$ & $0.51_{-0.51}^{+1.12}$ & $<1.20$ & $2.58_{-0.75}^{+0.94}$ \\
X19 & $6.34_{-1.22}^{+1.39}$ & $4.42_{-2.40}^{+3.54}$ & $9.62_{-3.47}^{+4.48}$ & $4.65_{-2.41}^{+3.63}$ & $2.38_{-1.12}^{+1.52}$ & $2.02_{-1.25}^{+1.85}$ & $1.62_{-1.26}^{+2.05}$ & $2.81_{-1.74}^{+2.88}$ & $3.49_{-1.47}^{+2.00}$ & $1.27_{-0.90}^{+1.44}$ & $2.59_{-1.50}^{+2.19}$ & $6.93_{-1.28}^{+1.46}$ \\
X20 & $2.22_{-0.63}^{+0.75}$ & $1.33_{-1.10}^{+2.28}$ & $1.26_{-0.99}^{+2.06}$ & $2.98_{-1.74}^{+2.82}$ & $3.69_{-1.27}^{+1.65}$ & $1.56_{-0.94}^{+1.56}$ & $7.13_{-2.41}^{+3.12}$ & $3.74_{-2.14}^{+3.15}$ & $1.50_{-1.05}^{+1.52}$ & $2.50_{-1.27}^{+1.80}$ & $0.40_{-0.40}^{+1.56}$ & $2.90_{-0.95}^{+1.14}$ \\
X21 & $1.92_{-0.55}^{+0.67}$ & $1.07_{-1.07}^{+2.33}$ & $4.06_{-2.09}^{+3.09}$ & $1.35_{-1.17}^{+2.39}$ & $18.40_{-3.13}^{+3.49}$ & $8.49_{-2.46}^{+3.06}$ & $7.49_{-2.53}^{+3.28}$ & $3.89_{-2.02}^{+3.04}$ & $11.00_{-2.59}^{+3.10}$ & $4.12_{-1.59}^{+2.10}$ & $13.40_{-3.29}^{+3.91}$ & $0.84_{-0.46}^{+0.65}$ \\
X22 & $1.35_{-0.47}^{+0.59}$ & $1.23_{-1.23}^{+2.48}$ & $1.14_{-1.14}^{+2.33}$ & $4.90_{-2.67}^{+4.03}$ & $5.76_{-1.79}^{+2.20}$ & $4.44_{-1.82}^{+2.47}$ & $<1.35$ & $1.24_{-0.95}^{+1.97}$ & $2.00_{-1.01}^{+1.50}$ & $2.07_{-1.03}^{+1.53}$ & $4.03_{-1.77}^{+2.47}$ & $1.71_{-0.62}^{+0.80}$ \\
X23 & $2.49_{-0.64}^{+0.77}$ & $3.02_{-1.79}^{+2.93}$ & $6.93_{-2.74}^{+3.72}$ & $4.23_{-2.29}^{+3.38}$ & $2.26_{-1.10}^{+1.49}$ & $6.49_{-2.19}^{+2.75}$ & $3.30_{-1.66}^{+2.50}$ & $4.16_{-2.07}^{+3.04}$ & $1.29_{-0.78}^{+1.29}$ & $1.97_{-1.02}^{+1.54}$ & $0.78_{-0.68}^{+1.39}$ & $1.72_{-0.63}^{+0.82}$ \\
X24 & $1.22_{-0.44}^{+0.57}$ & $1.43_{-1.10}^{+2.24}$ & $<1.67$ & -- & $2.69_{-1.11}^{+1.50}$ & $0.64_{-0.61}^{+1.23}$ & $<1.25$ & $2.60_{-1.54}^{+2.50}$ & $0.57_{-0.50}^{+1.02}$ & $0.59_{-0.51}^{+1.04}$ & $0.84_{-0.70}^{+1.45}$ & $0.17_{-0.17}^{+0.38}$ \\
X25 & $0.74_{-0.34}^{+0.48}$ & $<1.67$ & $<1.51$ & $<1.70$ & $<0.60$ & $<0.89$ & $<1.24$ & $<1.67$ & $0.32_{-0.32}^{+1.09}$ & $0.17_{-0.17}^{+1.21}$ & $<1.15$ & $<0.32$ \\
X26 & $1.21_{-0.44}^{+0.57}$ & -- & $<1.62$ & $<1.84$ & $<0.62$ & $0.61_{-0.59}^{+1.18}$ & $<1.30$ & $1.29_{-1.02}^{+2.11}$ & $4.04_{-1.59}^{+2.10}$ & $5.91_{-1.88}^{+2.40}$ & $3.79_{-1.67}^{+2.38}$ & $1.22_{-0.51}^{+0.70}$ \\
X27 & $0.62_{-0.32}^{+0.45}$ & $<1.73$ & $<1.56$ & $<1.73$ & $<0.65$ & $<0.95$ & $<1.26$ & -- & $<0.82$ & $<0.84$ & $<1.14$ & $0.42_{-0.33}^{+0.53}$ \\
X28 & $0.60_{-0.33}^{+0.45}$ & $6.11_{-2.68}^{+3.75}$ & $2.72_{-1.61}^{+2.62}$ & $2.97_{-1.84}^{+3.04}$ & $8.14_{-1.96}^{+2.35}$ & $3.88_{-1.57}^{+2.17}$ & $4.17_{-1.92}^{+2.74}$ & $2.76_{-1.75}^{+2.86}$ & $4.61_{-1.76}^{+2.32}$ & $5.65_{-1.92}^{+2.47}$ & $2.07_{-1.28}^{+2.12}$ & $5.58_{-1.16}^{+1.36}$ \\
X29 & $<0.20$ & $2.81_{-1.74}^{+2.87}$ & $3.84_{-1.99}^{+3.00}$ & $2.98_{-1.77}^{+2.87}$ & $3.60_{-1.29}^{+1.66}$ & $4.81_{-1.75}^{+2.33}$ & $4.20_{-1.85}^{+2.63}$ & $21.50_{-5.04}^{+6.03}$ & $13.60_{-2.87}^{+3.37}$ & $19.50_{-3.51}^{+4.00}$ & $24.40_{-4.50}^{+5.22}$ & $7.49_{-1.37}^{+1.52}$ \\
X30 & $0.13_{-0.13}^{+0.26}$ & -- & $<1.77$ & $1.52_{-1.26}^{+2.60}$ & $0.22_{-0.22}^{+0.99}$ & $0.70_{-0.67}^{+1.36}$ & $<1.43$ & $1.27_{-0.97}^{+2.01}$ & $3.32_{-1.35}^{+1.83}$ & $1.31_{-0.81}^{+1.34}$ & $4.66_{-1.88}^{+2.55}$ & $2.32_{-0.71}^{+0.89}$ \\
X31 & $0.10_{-0.10}^{+0.28}$ & $1.58_{-1.25}^{+2.58}$ & $<1.80$ & $<1.99$ & $0.24_{-0.24}^{+0.96}$ & $0.62_{-0.62}^{+1.35}$ & $<1.41$ & $<1.61$ & $<0.80$ & $<0.82$ & $<1.10$ & $0.69_{-0.39}^{+0.58}$ \\
X32 & $<0.20$ & $40.20_{-7.50}^{+8.54}$ & $29.10_{-5.96}^{+6.86}$ & $26.60_{-6.08}^{+7.13}$ & $86.80_{-6.95}^{+7.02}$ & $43.80_{-6.07}^{+6.14}$ & $49.10_{-7.50}^{+7.57}$ & $42.90_{-7.27}^{+8.20}$ & $45.70_{-5.65}^{+5.71}$ & $31.10_{-4.74}^{+4.79}$ & $53.40_{-7.15}^{+7.22}$ & $146.00_{-6.15}^{+6.21}$ \\
X33 & $<0.21$ & $1.39_{-1.15}^{+2.37}$ & $<1.69$ & $<1.84$ & $<0.69$ & $<1.03$ & $<1.42$ & $1.24_{-1.03}^{+2.12}$ & $1.26_{-0.80}^{+1.30}$ & $0.57_{-0.52}^{+1.05}$ & $0.81_{-0.70}^{+1.44}$ & $1.21_{-0.51}^{+0.69}$ \\
X34 & $0.39_{-0.27}^{+0.40}$ & $<1.79$ & -- & $1.40_{-1.11}^{+2.29}$ & $<0.65$ & $0.72_{-0.59}^{+1.23}$ & $1.00_{-0.76}^{+1.58}$ & $0.91_{-0.91}^{+2.13}$ & $<1.23$ & $0.77_{-0.77}^{+1.24}$ & $1.57_{-1.10}^{+1.81}$ & $3.50_{-1.02}^{+1.19}$ \\
X35 & $<0.21$ & -- & $<1.57$ & -- & $<0.65$ & -- & -- & $1.40_{-1.22}^{+2.48}$ & $6.17_{-2.04}^{+2.61}$ & $9.68_{-2.66}^{+3.18}$ & $7.38_{-2.63}^{+3.47}$ & $3.93_{-0.98}^{+1.17}$ \\
X36 & $0.97_{-0.40}^{+0.52}$ & $1.15_{-1.11}^{+2.21}$ & $2.36_{-1.59}^{+2.58}$ & $1.10_{-1.10}^{+2.25}$ & $1.74_{-0.96}^{+1.33}$ & $3.64_{-1.66}^{+2.25}$ & $1.91_{-1.32}^{+2.14}$ & $2.73_{-1.62}^{+2.63}$ & $1.23_{-0.78}^{+1.28}$ & $0.59_{-0.51}^{+1.04}$ & $1.71_{-1.08}^{+1.77}$ & $1.44_{-0.54}^{+0.71}$ \\
X37 & $<0.21$ & $<1.75$ & -- & $<1.78$ & $0.40_{-0.40}^{+0.80}$ & $<0.93$ & $<1.38$ & $2.56_{-1.63}^{+2.67}$ & $<0.85$ & $<0.87$ & $<1.13$ & $0.65_{-0.39}^{+0.58}$ \\
X38 & $<0.21$ & $4.39_{-2.18}^{+3.23}$ & $2.55_{-1.49}^{+2.41}$ & $7.42_{-2.92}^{+3.97}$ & $6.89_{-1.80}^{+2.15}$ & $4.00_{-1.63}^{+2.20}$ & $2.01_{-1.24}^{+2.05}$ & $1.34_{-1.02}^{+2.14}$ & $<0.84$ & $<0.85$ & $<1.17$ & $0.16_{-0.16}^{+0.38}$ \\
X39 & $0.28_{-0.20}^{+0.33}$ & $<1.73$ & $<1.51$ & $<1.72$ & $1.54_{-0.75}^{+1.13}$ & $<0.89$ & $<1.29$ & $1.26_{-1.06}^{+2.15}$ & $<0.85$ & $<0.88$ & $2.74_{-1.49}^{+2.20}$ & $<0.33$ \\
X40 & $<0.27$ & $<1.85$ & $<1.69$ & $4.48_{-2.32}^{+3.50}$ & $2.72_{-1.09}^{+1.50}$ & $1.70_{-1.01}^{+1.64}$ & $<1.31$ & $<1.71$ & $0.97_{-0.80}^{+1.28}$ & $0.30_{-0.30}^{+1.17}$ & $<1.19$ & $<0.51$ \\
X41 & $<0.21$ & -- & $1.23_{-0.93}^{+1.93}$ & -- & $<0.64$ & $1.63_{-0.95}^{+1.55}$ & -- & $<1.68$ & $<0.86$ & $0.31_{-0.31}^{+1.19}$ & $<1.25$ & $<0.31$ \\
X42 & $<0.22$ & $5.69_{-2.50}^{+3.49}$ & -- & $8.51_{-3.08}^{+4.07}$ & $1.01_{-0.60}^{+0.98}$ & $3.04_{-1.38}^{+1.93}$ & $0.93_{-0.81}^{+1.65}$ & $1.02_{-1.02}^{+2.09}$ & $2.06_{-1.07}^{+1.61}$ & $0.40_{-0.40}^{+1.12}$ & $1.74_{-1.14}^{+1.87}$ & $0.34_{-0.31}^{+0.48}$ \\
\hline
\end{tabular}
\tablefoot{
X-ray luminosities are given in the 0.3--10~keV band in units of $10^{38}$ erg s$^{-1}$.
The quoted lower and upper values are the luminosity uncertainties.
For non-detections with a valid upper limit, a ``$<$'' value is reported,
while ``--'' denotes epochs without a useful luminosity constraint.
}
\end{sidewaystable*}

\subsection{X-ray luminosity distributions and X-ray luminosity functions}
\label{sec:xlf}

Figure~\ref{fig:lx_hist} presents histograms of the 0.3–10 keV luminosity distribution of the X-ray sources in NGC~1300 for the 12 individual \textit{Chandra} observations, together with the corresponding upper limits for non-detections. Across the observations, the detected-source distributions are predominantly concentrated in the $10^{38}$--$10^{39}~{\rm erg~s^{-1}}$ range, although both the number of detected sources and the extent of the luminosity distribution vary from epoch to epoch. The most pronounced changes occur at the bright end, where several observations contain sources reaching or exceeding the ULX threshold of $10^{39}~{\rm erg~s^{-1}}$, while no source reaches this luminosity in ObsID 11775. The population above $10^{39}~{\rm erg~s^{-1}}$ is therefore not persistent across all epochs, reflecting the long-term luminosity variations of the brightest sources in the galaxy. At lower luminosities, the number of detected sources also changes appreciably between observations. However, the upper-limit distributions show corresponding epoch-to-epoch differences, demonstrating that the observations do not provide uniform sensitivity below $\sim10^{38}~{\rm erg~s^{-1}}$. The apparent variations in the faint-source population therefore cannot be interpreted independently of these sensitivity differences. In contrast, the recurrent concentration of detections between $10^{38}$ and $10^{39}~{\rm erg~s^{-1}}$ is a common feature of the individual observations. Taken together, the luminosity distributions show that the observed X-ray population is dominated by sources in this luminosity interval, while the principal temporal changes are associated with the most luminous sources and the appearance or disappearance of sources in the ULX regime.

\begin{figure*}
\begin{center}
\includegraphics[angle=0,scale=0.35]{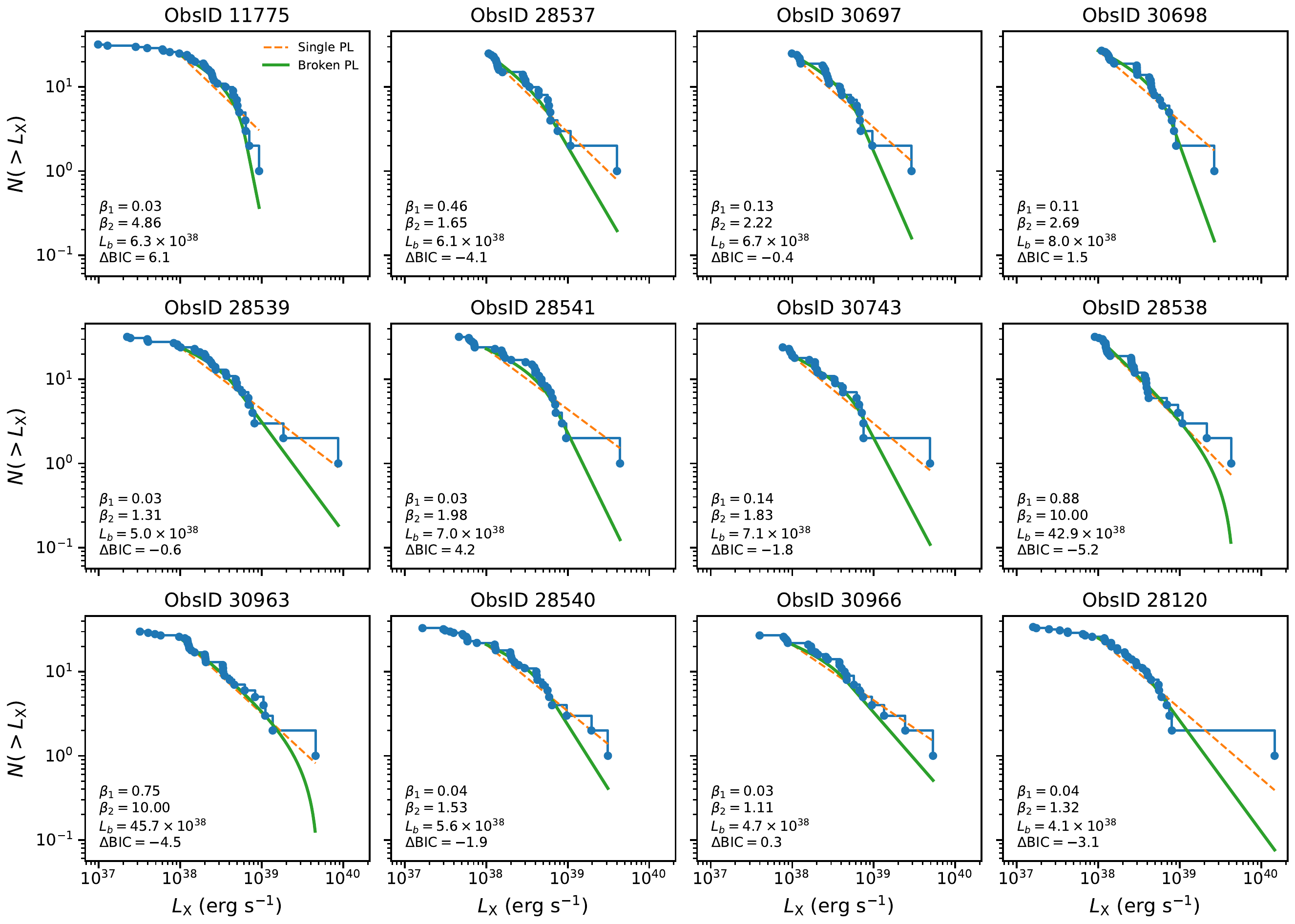}
\caption{Cumulative 0.3--10~keV X-ray luminosity functions (XLFs) of the off-nuclear sources in NGC~1300 for the 12 individual \textit{Chandra} observations. The nuclear source X12 is excluded. The XLFs are fitted above $L_{\rm X}=10^{38}$~erg~s$^{-1}$ with single and broken power-law models, shown by the dashed and solid curves, respectively. For the broken power-law fits, $\beta_1$ and $\beta_2$ denote the cumulative slopes below and above the break luminosity $L_{\rm b}$. The reported $\Delta{\rm BIC}$ is defined as ${\rm BIC}_{\rm single}-{\rm BIC}_{\rm broken}$, such that positive values favor the broken power-law model.}
\label{fig:lxf12}
\end{center}
\end{figure*}

Figure~\ref{fig:lxf12} shows the cumulative 0.3--10 keV X-ray luminosity functions (XLFs) constructed separately for the 12 \textit{Chandra} observations, excluding the nuclear source X12. The number of detected off-nuclear sources varies between epochs, and the XLFs extend from approximately $10^{37}$ erg s$^{-1}$ to above $10^{39}$ erg s$^{-1}$. To characterize the shape of the XLFs, we fitted the distributions above $L_{\rm X}=10^{38}$ erg s$^{-1}$ with a single power-law model, $N(>L_{\rm X})\propto L_{\rm X}^{-\beta}$. The resulting cumulative slopes span $\beta\simeq0.66$--$1.00$, indicating moderate epoch-to-epoch variations in the shape of the observed XLF. The flattest distribution is obtained for ObsID 30966 ($\beta\simeq0.66$), while the steepest is found for ObsID 28538 ($\beta\simeq1.00$). Despite these variations, the XLFs are generally well described by a single power law over the fitted luminosity range.

We additionally tested whether a broken power-law model provides a better description of the individual XLFs. For most observations, the Bayesian information criterion (BIC) does not favor the additional complexity of the broken power-law model. The strongest preference for a broken power law is found for ObsID 11775 ($\Delta{\rm BIC}=6.1$), with a break at $L_{\rm b}\simeq6.3\times10^{38}$ erg s$^{-1}$. ObsID 28541 shows weaker evidence ($\Delta{\rm BIC}=4.2$), with $L_{\rm b}\simeq7.0\times10^{38}$ erg s$^{-1}$. The remaining epochs either favor the single power-law description or provide only inconclusive evidence for a break. Since these XLFs are based on the observed source distributions and have not been corrected for incompleteness, the apparent flattening and breaks at the low-luminosity end cannot be interpreted as intrinsic features of the NGC~1300 source population.

\subsection{Spectral hardness and X-ray colors}
\label{sec:hardness}

\begin{figure*}
\begin{center}
\includegraphics[angle=0,scale=0.4]{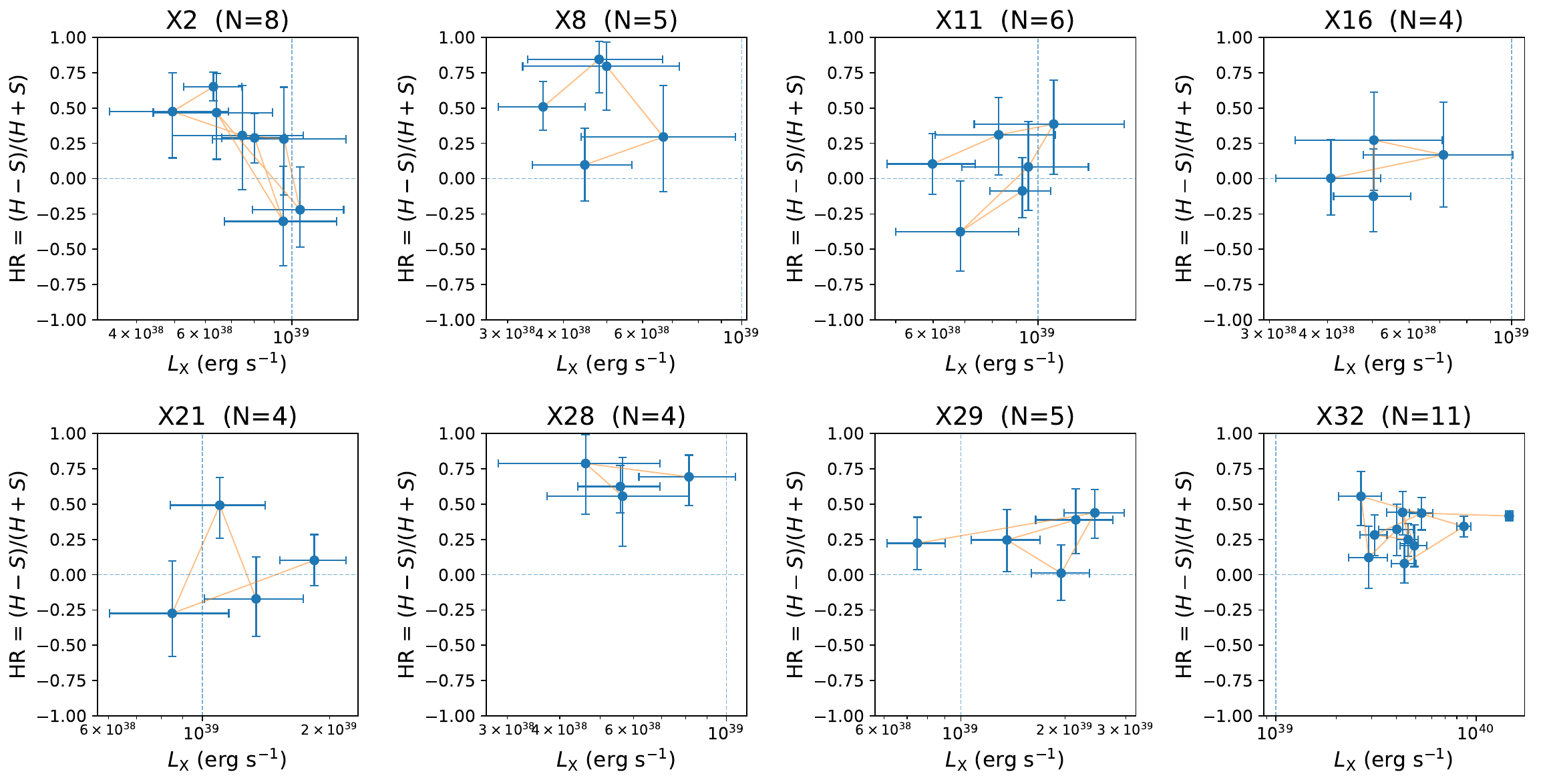}
\caption{Hardness--intensity diagrams for the selected ULXs and ULX candidates in NGC~1300. The hardness ratio is defined as ${\rm HR}=(H-S)/(H+S)$, where $S$ and $H$ are the 0.3--2 and 2--10~keV fluxes, respectively, while the intensity is represented by the 0.3--10~keV X-ray luminosity, $L_{\rm X}$. The sample comprises five ULXs (X2, X11, X21, X29, and X32) and three ULX candidates (X8, X16, and X28). Only epochs for which both the lower and upper uncertainties on HR are $\leq 0.40$ are shown, and each source is required to have at least four such measurements. Error bars represent the propagated uncertainties in HR and $L_{\rm X}$. The horizontal dashed line marks ${\rm HR}=0$, and the vertical dashed line indicates the ULX luminosity threshold at $L_{\rm X}=10^{39}~{\rm erg~s^{-1}}$. The orange lines connect consecutive epochs in chronological order.} 
\label{fig:hid}
\end{center}
\end{figure*}

Figure~\ref{fig:hid} shows the hardness--intensity diagrams of the eight ULXs and ULX candidates with sufficiently well-constrained hardness measurements. X8 and X28 remain predominantly hard throughout the sampled luminosity range, with all selected measurements at ${\rm HR}>0$. X16 also shows comparatively modest hardness variations, although only four epochs satisfy the selection criteria. In contrast, X2, X11, and X21 exhibit stronger epoch-to-epoch changes in hardness. X2 spans ${\rm HR}\simeq-0.3$--$0.7$ and reaches its softest values at the highest sampled luminosities. X11 and X21 both cross ${\rm HR}=0$, indicating substantial changes in their broad-band spectral hardness between observations. X29 also shows measurable hardness variations while spanning approximately an order of magnitude in luminosity, but no clear monotonic relation between hardness and luminosity is apparent. X32 displays a distinct behavior. Its luminosity varies over a broad range, extending from a few $10^{39}$ to above $10^{40}~{\rm erg~s^{-1}}$, whereas its hardness remains comparatively stable and predominantly positive. Most measurements are concentrated around ${\rm HR}\sim0.1$--$0.5$, and no systematic hardening or softening with increasing luminosity is evident. Thus, the pronounced long-term luminosity variability of X32 is not accompanied by a comparably large change in its broad-band X-ray hardness.

To further characterize X32, we examined its position in the X-ray color--color diagram using the classification scheme of \citet{2003ApJ...595..719P}. Figure~\ref{fig:colorcolor} shows the five epochs for which the source is constrained in all three energy bands. The measurements occupy a relatively narrow region of the diagram, with soft colors of approximately $0.13$--$0.38$ and hard colors of approximately $0.16$--$0.40$. All five measurements have positive soft and hard colors and overlap the region associated with HMXBs in the X-ray color--color classification of \citet{2003ApJ...595..719P}. This provides additional evidence that the X-ray properties of X32 are consistent with an HMXB-like source, although X-ray colors alone cannot uniquely determine the nature of an individual binary system. Moreover, for X32, the available photon statistics allowed spectral fitting in ObsIDs~28120 and 28539 (Table~\ref{tab:x32_spec}; Fig.~\ref{fig:modelsX32}). For ObsID~28120, the absorbed \textit{diskbb} and power-law fits yield C-statistic values of 64.75 and 76.86, respectively, over the original fitting band. Given the residuals near the band edges (Fig.~\ref{fig:modelsX32}), we repeated both fits over 0.5--6.0~keV. In this band, the \textit{diskbb} and power-law fits yield C-statistic values of 62.41 and 67.47, respectively, for the same 58 spectral bins. The \textit{diskbb} model still yields a lower C-statistic, but the difference decreases from $\Delta C=12.11$ to $5.06$.

In ObsID~28120, the \textit{diskbb} model provides a noticeably better fit than the power-law model, favoring a disk-dominated spectral description. In ObsID~28539, the two models provide statistically comparable fits, although the spectral parameters are less well constrained because of the lower photon statistics. The remaining observations of X32, as well as those of the other ULXs and ULX candidates, do not contain sufficient counts to meaningfully constrain their spectral parameters.

\begin{table}
\caption{Best-fitting spectral parameters of X32 for two
\textit{Chandra} observations.}
\label{tab:x32_spec}
\centering
\begin{tabular}{lcc}
\hline\hline
Parameter & \textit{powerlaw} & \textit{diskbb} \\ \hline \multicolumn{3}{c}{ObsID 28120} \\
\hline $N_{\rm H}$ ($10^{22}$ cm$^{-2}$) & $0.82_{-0.29}^{+0.31}$ & $0.28_{-0.19}^{+0.21}$ \\ $\Gamma$ & $2.24_{-0.29}^{+0.30}$ & -- \\ 
$kT_{\rm in}$ (keV) & -- & $1.37_{-0.16}^{+0.21}$ \\ 
Norm. & $1.07_{-0.34}^{+0.52}\times10^{-4}$ & $3.56_{-1.60}^{+2.69}\times10^{-3}$ \\ 
$F_{\rm X}^{\rm abs}$ & $2.26_{-0.08}^{+0.11}\times10^{-13}$ & $2.03_{-0.11}^{+0.16}\times10^{-13}$ \\
$L_{\rm X}^{\rm unabs}$ &  $2.78_{-0.08}^{+0.10}\times10^{40}$ &  $1.30_{-0.10}^{+0.18}\times10^{40}$ \\ 
C-stat/d.o.f. & $76.86/56$ & $64.75/56$ \\ 
\hline \multicolumn{3}{c}{ObsID 28539} \\ 
\hline

$N_{\rm H}$ ($10^{22}$ cm$^{-2}$) &
$0.77_{-0.68}^{+0.73}$ &
$0.25_{-0.25}^{+0.49}$ \\

$\Gamma$ &
$2.02_{-0.66}^{+0.69}$ &
-- \\

$kT_{\rm in}$ (keV) &
-- &
$1.52_{-0.39}^{+0.81}$ \\

Norm. &
$5.57_{-3.27}^{+8.11}\times10^{-5}$ &
$1.54_{-1.54}^{+4.12}\times10^{-3}$ \\

$F_{\rm X}^{\rm abs}$ &
$1.60_{-0.14}^{+0.16}\times10^{-13}$ &
$1.40_{-0.06}^{+0.11}\times10^{-13}$ \\
$L_{\rm X}^{\rm unabs}$ &
$1.59_{-0.08}^{+0.12}\times10^{40}$ &
$8.57_{-0.06}^{+0.10}\times10^{39}$ \\

C-stat/d.o.f. &
$21.85/16$ &
$21.47/16$ \\
\hline
\end{tabular}
\tablefoot{ $N_{\rm H}$ is the equivalent hydrogen column density, $\Gamma$ the power-law photon index, $kT_{\rm in}$ the inner-disk temperature, and Norm. the model normalization. $F_{\rm X}^{\rm abs}$ denote the absorbed 0.3--10~keV fluxes, and $L_{\rm X}^{\rm unabs}$ is the corresponding unabsorbed luminosity assuming a distance of 20.7~Mpc. Parameter uncertainties correspond to the 90\% confidence level ($\Delta C=2.706$ for one parameter of interest). For ObsID~28539, the lower confidence bounds of $N_{\rm H}$ and the \textit{diskbb} normalization reached their hard limits and are therefore unconstrained toward lower values.}
\end{table}

\section{Discussion}

\subsection{Long-term variability and the ULX population}

The multi-epoch \textit{Chandra} observations reveal a luminous and highly dynamic X-ray source population in NGC~1300. The luminosity classification and long-term variability properties of the individual sources are summarized in Table~\ref{tab:source_classification}. Excluding the nuclear source X12, 41 sources are detected within the galaxy, with individual sources showing markedly different long-term behavior over the $\sim16$ yr baseline. The majority of the population reaches luminosities above $10^{38}~{\rm erg~s^{-1}}$ in at least one observation, while six sources (X2, X11, X18, X21, X29, and X32) are observed above the conventional ULX threshold of $10^{39}~{\rm erg~s^{-1}}$. A further 13 sources have luminosity uncertainties extending into the ultraluminous regime and are therefore classified as ULX candidates, while the remaining off-nuclear population comprises 20 bright and two faint sources. The resulting population is consistent with the expectation that luminous X-ray binaries in star-forming galaxies are dominated by HMXBs, whose numbers and luminosity distribution are closely linked to recent star formation \citep{2003MNRAS.339..793G,2006ARA&A..44..323F, 2012MNRAS.419.2095M}. The repeated observations also demonstrate, however, that a luminosity-based census obtained from a single epoch would provide an incomplete description of the population. The number of sources above $10^{39}~{\rm erg~s^{-1}}$ changes substantially between observations, and several objects move between luminosity regimes during the monitoring campaign. Thus, the distinction between bright X-ray binaries and ULXs in NGC~1300 is not persistent for a significant subset of the population, but depends on the epoch at which the galaxy is observed.

The observed long-term variability further supports this interpretation. Ten sources (X7, X9, X13, X29, X32, X35, X37, X38, X40, and X42) satisfy our operational transient criterion, showing at least one non-detection together with a ratio of at least 10 between the peak luminosity and the minimum luminosity upper limit among the non-detections. Seven sources (X7, X12, X13, X21, X28, X29, and X32) also show significant long-term variability with $S_{\rm var}\geq3$. Four of these sources (X7, X13, X29, and X32) satisfy both criteria and are therefore both transient and significantly variable according to our adopted definitions. Among the ten transient sources, two are secure ULXs (X29 and X32), four are ULX candidates (X9, X35, X38, and X42), and four are bright sources (X7, X13, X37, and X40). The largest dynamic ranges occur among some of the most luminous objects: X29 varies by at least a factor of 119 and X32 by more than a factor of 700. Large-amplitude variability and transient ultraluminous episodes are now observed in a growing number of extragalactic X-ray binaries \citep{2019MNRAS.483.3566V,2020ApJ...891..153E, 2023ApJ...951...51B,2023MNRAS.526.5765A}. 

In particular, \citet{2023ApJ...951...51B} showed that repeated monitoring can reveal transient ULXs that would be absent from a luminosity-selected sample during their faint states. The NGC~1300 population illustrates the same observational bias: several sources are detected only during a fraction of the available epochs, and some enter the ultraluminous regime only temporarily. Such behavior can reflect genuine changes in the accretion rate, disc-instability-driven outbursts, or changes in the observed emission associated with super-Eddington accretion geometry. Disc-instability models can in particular produce transient ULX phases under appropriate binary and mass-transfer conditions \citep{2020A&A...643A.171H}, whereas for neutron-star accretors large luminosity changes may additionally be associated with transitions between different magnetospheric accretion regimes. The present sampling does not allow these mechanisms to be distinguished for the individual sources, but it clearly establishes long-term variability as a major property of the NGC~1300 X-ray population.

\begin{figure}
\resizebox{\hsize}{!}{\includegraphics{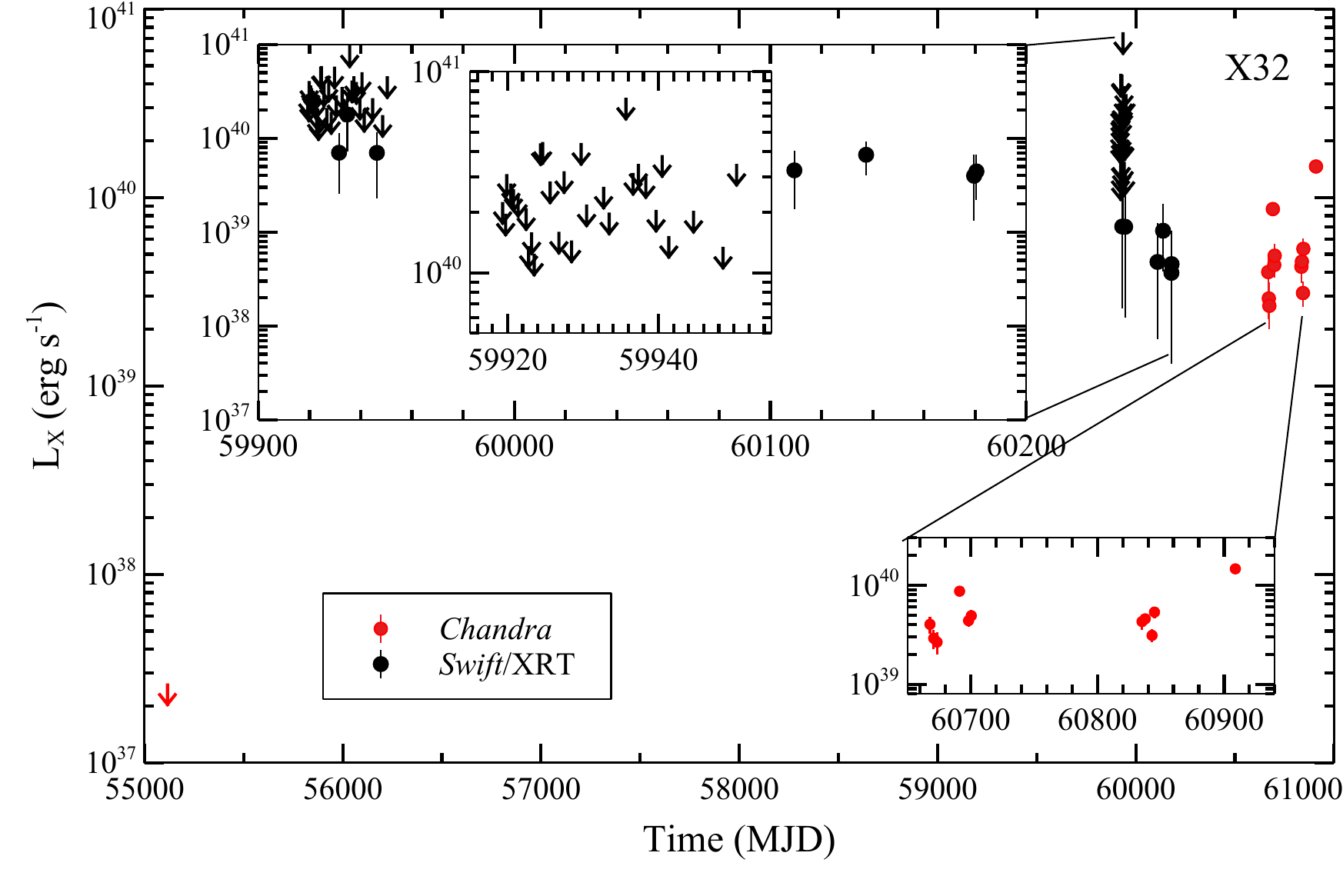}}
\caption{Long-term 0.3--10~keV X-ray luminosity evolution of X32 in NGC~1300. Red and black points show the \textit{Chandra} and \textit{Swift}/XRT X-ray luminosities, respectively, and downward arrows indicate upper limits. The upper inset shows an enlarged view of the \textit{Swift}/XRT observations, with a further zoom on the densely sampled interval around MJD~59920--59950. The lower inset shows an enlarged view of the \textit{Chandra} observations obtained in 2024--2025.}
\label{F:lcx32}
\end{figure}

The ULX population itself displays considerable diversity rather than a single characteristic variability pattern. X2, X11, X18, and X21 are detected throughout, or nearly throughout, the available observations and repeatedly reach high luminosities, whereas X29 and X32 combine ultraluminous peak luminosities with much larger dynamic ranges. X32 is the most extreme source in the galaxy, reaching $L_{\rm X,max}\simeq1.46\times10^{40}~{\rm erg~s^{-1}}$, while X29 reaches $\simeq2.44\times10^{39}~{\rm erg~s^{-1}}$ (see Fig.~\ref{F:lcx32}). The coexistence of comparatively persistent and strongly transient ULXs is qualitatively consistent with the heterogeneous phenomenology of the ULX population, which is now known to contain both black-hole and neutron-star accretors operating in the super-Eddington regime \citep{2013MNRAS.435.1758S,2018ApJ...856..128W, 2020MNRAS.494.3611K}.

The 13 additional ULX candidates are also important in this context. Their luminosity confidence intervals reach $10^{39}~{\rm erg~s^{-1}}$, and four of them---X9, X35, X38, and X42---are transient according to our variability criterion. They may therefore represent systems observed close to, or only intermittently entering, the ultraluminous regime rather than a physically separate source population. Consequently, the six secure ULXs should be regarded as the conservative observed ULX population of NGC~1300, while the candidates trace a broader population of luminous and potentially super-Eddington X-ray binaries. Distinguishing between these possibilities will require deeper spectra and denser temporal coverage, particularly for the transient candidates, because luminosity and variability alone cannot uniquely determine either the nature of the compact object or the accretion geometry.

\begin{table*}
\centering
\caption{Luminosity and long-term variability properties of the
X-ray sources in NGC~1300.}
\label{tab:source_classification}
\setlength{\tabcolsep}{4.0pt}
\small
\begin{tabular}{cccccccc}
\hline\hline
Source &
ObsID$_{\rm max}$ &
$L_{\rm X,max}$ &
$N_{\rm det}/12$ &
Luminosity class &
$R$ &
$S_{\rm var}$ &
Variability class \\
&
&
($10^{39}$ erg s$^{-1}$) &
&
&
&
&
\\
\hline
X1 & 30697 & 0.27 & 11/12 & Bright & $>2.3$ & 0.96 & No significant \\
X2 & 30963 & 1.05 & 12/12 & ULX & $4.1$ & 2.17 & No significant \\
X3 & 28539 & 0.20 & 9/12 & Bright & $>2.4$ & 1.10 & No significant \\
X4 & 28540 & 0.22 & 8/12 & Bright & $>2.4$ & 1.58 & No significant \\
X5 & 30697 & 0.68 & 12/12 & ULX candidate & $7.4$ & 1.87 & No significant \\
X6 & 30698 & 0.57 & 10/12 & Bright & $>6.7$ & 1.84 & No significant \\
X7 & 11775 & 0.44 & 3/12 & Bright & $>14.6$ & 3.00 & Transient \\
X8 & 30698 & 0.90 & 11/12 & ULX candidate & $3.7$ & 1.58 & No significant \\
X9 & 30697 & 0.67 & 6/12 & ULX candidate & $>21.5$ & 1.53 & Transient \\
X10 & 28540 & 0.43 & 9/12 & Bright & $>3.7$ & 2.23 & No significant \\
X11 & 28538 & 1.08 & 12/12 & ULX & $5.3$ & 2.31 & No significant \\
X12 & 30698 & 5.72 & 12/12 & Nuclear & $3.1$ & 3.98 & Variable \\
X13 & 11775 & 0.45 & 2/12 & Bright & $>15.1$ & 3.05 & Transient \\
X14 & 30697 & 0.62 & 12/12 & ULX candidate & $7.5$ & 1.82 & No significant \\
X15 & 28537 & 0.31 & 9/12 & Bright & $>3.8$ & 1.13 & No significant \\
X16 & 28541 & 0.71 & 12/12 & ULX candidate & $5.8$ & 2.21 & No significant \\
X17 & 28539 & 0.77 & 12/12 & ULX candidate & $6.3$ & 2.21 & No significant \\
X18 & 28537 & 1.08 & 11/12 & ULX & $>9.0$ & 2.71 & No significant \\
X19 & 30697 & 0.96 & 12/12 & ULX candidate & $7.6$ & 2.22 & No significant \\
X20 & 30743 & 0.71 & 12/12 & ULX candidate & $17.9$ & 2.34 & No significant \\
X21 & 28539 & 1.84 & 12/12 & ULX & $21.8$ & 5.49 & Variable \\
X22 & 28539 & 0.58 & 11/12 & Bright & $>4.3$ & 1.57 & No significant \\
X23 & 30697 & 0.69 & 12/12 & ULX candidate & $8.9$ & 2.00 & No significant \\
X24 & 28539 & 0.27 & 9/12 & Bright & $>2.2$ & 2.15 & No significant \\
X25 & 11775 & 0.07 & 3/12 & Faint & $>2.3$ & 0.46 & No significant \\
X26 & 28540 & 0.59 & 7/12 & Bright & $>9.6$ & 2.39 & No significant \\
X27 & 11775 & 0.06 & 2/12 & Faint & $>1.0$ & 0.32 & No significant \\
X28 & 28539 & 0.81 & 12/12 & ULX candidate & $13.5$ & 3.75 & Variable \\
X29 & 30966 & 2.44 & 11/12 & ULX & $>119.0$ & 4.05 & Transient \\
X30 & 30966 & 0.47 & 9/12 & Bright & $>3.3$ & 2.39 & No significant \\
X31 & 28537 & 0.16 & 5/12 & Bright & $>2.0$ & 1.16 & No significant \\
X32 & 28120 & 14.60 & 11/12 & ULX & $>715.7$ & 12.68 & Transient \\
X33 & 28537 & 0.14 & 6/12 & Bright & $>6.5$ & 0.53 & No significant \\
X34 & 28120 & 0.35 & 8/12 & Bright & $>5.4$ & 2.84 & No significant \\
X35 & 28540 & 0.97 & 5/12 & ULX candidate & $>45.2$ & 2.28 & Transient \\
X36 & 28541 & 0.36 & 12/12 & Bright & $6.2$ & 1.56 & No significant \\
X37 & 28538 & 0.26 & 3/12 & Bright & $>12.4$ & 1.19 & Transient \\
X38 & 30698 & 0.74 & 8/12 & ULX candidate & $>34.7$ & 2.47 & Transient \\
X39 & 30966 & 0.27 & 4/12 & Bright & $>8.4$ & 1.61 & No significant \\
X40 & 30698 & 0.45 & 5/12 & Bright & $>16.9$ & 1.61 & Transient \\
X41 & 28541 & 0.16 & 3/12 & Bright & $>7.7$ & 0.87 & No significant \\
X42 & 30698 & 0.85 & 10/12 & ULX candidate & $>39.2$ & 2.62 & Transient \\
\hline
\end{tabular}

\begin{flushleft}
\footnotesize
\textit{Notes.}
$L_{\rm X,max}$ is the maximum measured 0.3--10~keV luminosity, and $N_{\rm det}$ is the number of detections in 12 \textit{Chandra} observations. Sources with $L_{\rm X,max}\geq10^{39}~{\rm erg~s^{-1}}$ are classified as ULXs, excluding the nuclear source X12. Sources below this threshold are classified as ULX candidates if the upper uncertainty bound reaches the threshold in at least one epoch. Significant variability is defined by $S_{\rm var}\geq3$. Transients have at least one non-detection and $R=L_{\rm X,max}/L_{\rm X,UL}\geq10$, where $L_{\rm X,UL}$ is the minimum luminosity upper limit among the non-detections. Transients can also satisfy the significant variability criterion. Sources satisfying neither criterion are listed as ``No significant.
\end{flushleft}
\end{table*}

\subsection{X-ray spectral variability}

The hardness--intensity diagrams indicate diverse spectral behavior among the luminous X-ray sources in NGC~1300. X8 and X28 remain predominantly hard over the sampled luminosity range, whereas X2, X11, and X21 show larger epoch-to-epoch changes in hardness, with several measurements extending across ${\rm HR}=0$. X29 also varies in both luminosity and hardness, but without a clear monotonic relation between the two quantities. These trends should, however, be interpreted with some caution, since the hardness ratios of most sources have relatively large uncertainties, particularly when one of the two energy bands is weakly detected. Indeed, only eight sources retained at least four epochs after restricting the analysis to measurements with HR uncertainties $\leq0.40$. The available data therefore do not support a unique hardness--luminosity trend for the population as a whole. This diversity is nevertheless broadly consistent with the heterogeneous spectral evolution observed in ULXs, where luminosity variations may be accompanied by different degrees of spectral change \citep{2009ApJ...696.1712F,2013MNRAS.435.1758S,2017ARA&A..55..303K}. Such behavior can arise from changes in the relative contributions of the inner accretion flow and an optically thick wind, together with viewing-angle effects in super-Eddington accretion flows \citep{2009MNRAS.397.1836G,2013MNRAS.435.1758S}.

X32 provides the clearest hardness evolution in our sample because its higher count statistics allow its hardness to be followed over 11 epochs with comparatively well-constrained measurements. Despite reaching $L_{\rm X}\simeq1.5\times10^{40}~{\rm erg~s^{-1}}$ and undergoing large long-term luminosity variations, X32 remains predominantly hard, with most measurements at ${\rm HR}\sim0.1$--$0.5$ and no clear systematic hardness--luminosity trend. Its color--color behavior independently supports this picture: all five epochs with constrained measurements in the 0.3--1, 1--2, and 2--8~keV bands lie within, or are consistent within their uncertainties with, the HMXB locus defined by \citep{2003ApJ...595..719P}. The clustering of these measurements in the same region of the color--color diagram, together with the persistently positive broad-band hardness ratios, supports an HMXB-like classification for X32 and indicates that its large luminosity changes are not accompanied by comparably strong changes in broad-band spectral shape. However, the classification of \citep{2003ApJ...595..719P} is empirical and based on broad-band colors, and therefore provides supporting evidence for an HMXB interpretation rather than a unique identification of the binary system.

X32 is particularly noteworthy because it is identified as a ULX for the first time in this work and exhibits a transient behavior with an extreme luminosity ratio of $R>715$. The spectral fits strengthen this classification, while also illustrating the model dependence of its intrinsic luminosity. In particular, the power-law fits favor a substantially higher $N_{\rm H}$ than the \textit{diskbb} fits, resulting in a stronger absorption correction and correspondingly higher unabsorbed luminosities. The luminosities obtained from the uniform \textit{srcflux} analysis are lower than those inferred from the power-law fits, as expected from the fixed $N_{\rm H}=2\times10^{20}$~cm$^{-2}$ adopted for the population analysis, but remain consistent with the source reaching the ULX regime. Thus, the classification of X32 as a ULX is not driven by a particular choice of spectral model. For ObsID~28120, the \textit{diskbb} fit yields a lower C-statistic than the power-law fit, although the difference decreases from $\Delta C=12.11$ to $5.06$ over 0.5--6.0~keV. Residuals near the band edges may therefore account for part of the difference between the two fits, and the evidence for intrinsic spectral curvature remains limited. Inner-disk color temperatures of $kT_{\rm in}\sim1$--$3$~keV, as inferred from phenomenological multicolor-disk fits, have been reported for stellar-remnant accretors at high accretion rates \citep{2009MNRAS.397.1836G}. Moreover, broadened disk-like spectra with $kT_{\rm in}\sim1$--$2$~keV have been identified with the broadened-disk ULX spectral regime, which is commonly interpreted as being associated with accretion near or above the Eddington limit \citep{2013MNRAS.435.1758S}. The \textit{diskbb} temperature should therefore not be interpreted as the temperature of a standard thin disk or used directly to infer the compact-object mass. The spectral properties and large luminosity variations of X32 are consistent with super-Eddington accretion. 

To further assess the nature of the compact object, we compared X32 with the hardness--softness diagram of \citet{2017ApJ...836..113P}. For ObsID~28120, the power-law model yields hardness and softness ratios of 4.09 and 1.53, respectively, while the \textit{diskbb} model gives 0.72 and 1.80. In their Fig.~2, the power-law prediction lies beyond the plotted hardness range, whereas the \textit{diskbb} prediction occupies a softer region than the confirmed ULX pulsars. The different predictions reflect the model dependence of the extrapolated 6--30~keV flux, preventing a reliable identification of the compact object as either a neutron star or a black hole.

\subsection{X-ray luminosity function and source population}

\begin{figure}
\resizebox{\hsize}{!}{\includegraphics{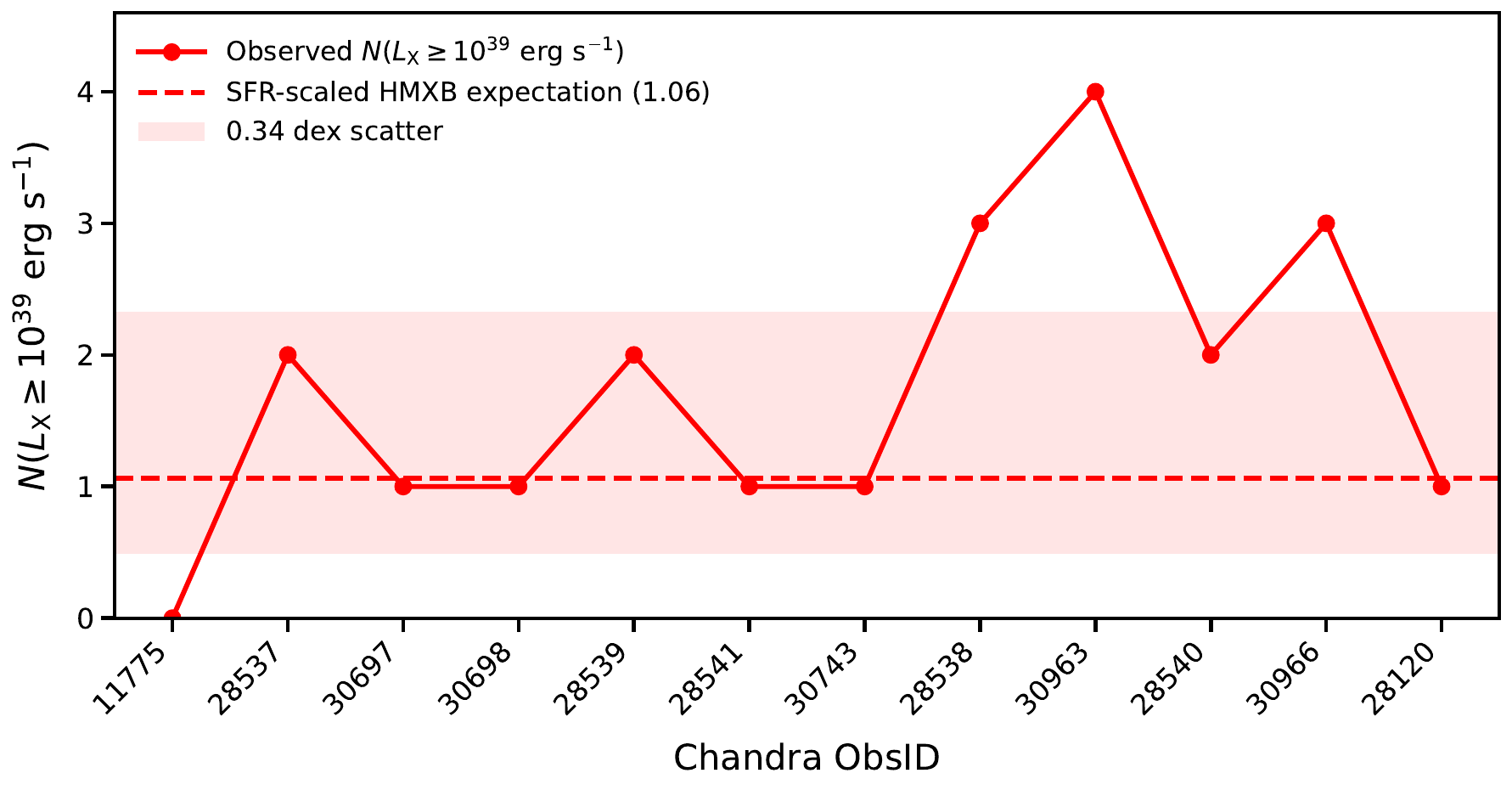}}
\caption{Number of sources with $L_{\rm X}\geq10^{39}~{\rm erg~s^{-1}}$ in each \textit{Chandra} observation of NGC~1300. The dashed line shows the number of HMXBs predicted from the SFR-scaled XLF of \citep{2012MNRAS.419.2095M} for ${\rm SFR}=1.4~M_\odot\,{\rm yr}^{-1}$, and the shaded region indicates the 0.34-dex scatter in the XLF normalization.} 
\label{fig:sfr_expected}
\end{figure}

The NGC 1300 XLF is broadly consistent with that expected for a luminous X-ray binary population, but its shape alone does not distinguish between HMXB- and LMXB-dominated populations. The cumulative slopes of $\beta\simeq0.66$--$1.00$ are comparable to the canonical HMXB XLF measured in nearby star-forming galaxies, for which a differential slope of $\sim1.6$ corresponds to a cumulative slope of $\sim0.6$ \citep{2012MNRAS.419.2095M}. The HMXB comparison can also be examined through the XLF normalization. For the adopted SFR of NGC~1300, $1.4~M_\odot\,{\rm yr}^{-1}$, the HMXB XLF of \citet{2012MNRAS.419.2095M} predicts 1.1 sources above $10^{39}~{\rm erg~s^{-1}}$. An independent estimate can be obtained from the empirical ULX--SFR scaling relation of \citet{2020MNRAS.498.4790K}, $N_{\rm ULX}\simeq0.44\,{\rm SFR}$, which predicts $\sim0.6$ ULXs for NGC~1300. The observed number of off-nuclear sources above $10^{39}~{\rm erg~s^{-1}}$ ranges from zero to four across the individual \textit{Chandra} epochs, reaching three in ObsIDs~28538 and 30966 and four in ObsID~30963 (see Fig.~\ref{fig:sfr_expected}). Thus, at some epochs the observed number of ULX-luminosity sources exceeds the nominal expectation based on the present-day SFR. However, given the intrinsic scatter in the HMXB XLF normalization, the small-number statistics, and the pronounced long-term variability of the brightest sources, this apparent excess does not by itself require an enhanced formation efficiency of luminous XRBs in NGC~1300. The XLF properties nevertheless support a substantial HMXB contribution, while a quantitative separation of the HMXB and LMXB components would require additional constraints from the spatial distributions of star formation and stellar mass.

The barred morphology of NGC~1300 may provide additional context for the spatial distribution of its luminous XRB population. In the barred spiral NGC~1672, \citet{2011ApJ...734...33J} found that the three brightest ULXs are located near the ends of the stellar bar, where gas accumulation and enhanced star formation may favor the formation of young XRB populations. This result suggests that the global SFR alone may not fully characterize the environments relevant to ULX production in barred galaxies. In NGC~1300, the pronounced variation in star-formation efficiency between the bar, spiral arms, and bar-end regions therefore motivates examining whether the luminous X-ray sources preferentially trace particular star-forming environments. The present-day SFR may not fully trace the stellar population from which the observed HMXBs originated. HMXB formation is delayed with respect to the onset of star formation, and both the number and luminosity distribution of HMXBs therefore depend on the recent star-formation history \citep{2016MNRAS.459..528A,2024ApJ...977..189L}. A higher level of star formation in the recent past could thus produce a larger present-day HMXB population than inferred from a scaling based solely on the current SFR. Whether the luminous XRB population of NGC~1300 retains such an imprint of its recent star-formation history cannot be established from the present X-ray data alone and requires spatially resolved constraints on the age distribution of its young stellar populations. The ULX population may also depend on metallicity. Observational studies show that the number of ULXs per unit SFR increases toward lower metallicities \citep{2013ApJ...769...92P,2020MNRAS.498.4790K}. Moreover, \citet{2019ApJS..243....3L} found an excess of luminous ($L_{\rm X}\gtrsim10^{39}~{\rm erg~s^{-1}}$) HMXBs in low-metallicity galaxies relative to the average HMXB XLF. NGC~1300 exhibits a negative radial metallicity gradient, with a central oxygen abundance of $12+\log({\rm O/H})\simeq8.66$, implying progressively lower abundances toward the outer disk. Such spatial metallicity variations may therefore contribute to differences in the production of luminous HMXBs across the galaxy. However, assessing this effect for the present sample requires metallicity measurements specifically at the locations of the X-ray sources.

The multi-epoch observations show that the bright end of the XLF is strongly affected by long-term source variability. X29 and X32 reach the ULX regime but span luminosity ranges of $R>119$ and $R>715$, respectively, while several ULX candidates also undergo large-amplitude changes. These sources can therefore enter or leave the bright tail between observations. Similar behavior has been observed in transient ULXs that fall below the luminosity-based ULX threshold during their faint states \citep{2020ApJ...891..153E, 2023ApJ...951...51B}. In NGC~1300, this effect is particularly important above $10^{39}~{\rm erg~s^{-1}}$, where only a few sources populate the cumulative distribution and a single outburst can substantially alter its bright end. The variation from zero to four ULX-luminosity sources between epochs can therefore be understood, at least in part, as a consequence of the duty cycles and long-term accretion variability of the brightest binaries rather than a change in the underlying XRB population.

Interpretation of the faint end is more strongly limited by observational selection and background contamination. The expected $\sim16$ background AGNs within the $D_{25}$ region, inferred from the $\log N$--$\log S$ relation of \citet{2009A&A...497..635C}, constitute a substantial statistical fraction of the off-nuclear catalog and should preferentially affect the lower-luminosity part of the XLF. Differences in sensitivity among the individual \textit{Chandra} observations introduce an additional luminosity-dependent incompleteness. These effects are particularly relevant to the tentative preference for a broken power law in some epochs, since incompleteness near the detection limit can produce an apparent flattening or change in slope. The lack of a consistently detected break across the 12 epochs therefore argues against associating the fitted break luminosities with a specific accretion transition or XRB population. The bright end is less affected by these limitations and more directly traces the luminous XRB population, although a completeness-corrected XLF together with spatially resolved star-formation and stellar-mass information would be required for quantitative constraints on the HMXB and LMXB components.

\section{Conclusion and summary}
\label{C}

We analyzed 12 archival \textit{Chandra} ACIS-S observations of NGC~1300 spanning more than 15 years, including 11 observations obtained in 2024--2025 that have not previously been used to characterize its X-ray source population. Source detection was performed independently in each observation. We detected 42 X-ray sources within the $D_{25}$ region, one of which (X12) is associated with the nucleus. Among the 41 off-nuclear sources, six (X2, X11, X18, X21, X29, and X32) reach the ULX regime, while 13 additional sources are classified as ULX candidates because their luminosity uncertainties extend to or above $10^{39}~{\rm erg~s^{-1}}$. The remaining population comprises 20 bright sources with $10^{38}\leq L_{\rm X}<10^{39}~{\rm erg~s^{-1}}$ and two faint sources with $L_{\rm X}<10^{38}~{\rm erg~s^{-1}}$. This substantially expands the previously known luminous X-ray source population of NGC~1300 and establishes its multi-epoch ULX census.

Long-term variability is a defining property of this population. Ten sources satisfy our transient criterion, including the ULXs X29 and X32 and four ULX candidates. Seven sources have $S_{\rm var}\geq3$, with X7, X13, X29, and X32 satisfying both variability criteria. X32 is the most extreme system, reaching $L_{\rm X,max}=1.46\times10^{40}~{\rm erg~s^{-1}}$ and $R>715$, and is identified as a transient ULX for the first time in this work. X29 is also a transient ULX, with $R>119$. In contrast, X2, X11, X18, and X21 are detected throughout or nearly throughout the observations. The NGC~1300 ULX population therefore contains both persistent luminous systems and sources that undergo large-amplitude long-term variability.

The spectral analysis further distinguishes X32 from the other luminous sources. Its broad-band hardness remains predominantly positive despite its extreme luminosity variations, and its X-ray colors overlap the empirical HMXB region. For ObsID~28120, the \textit{diskbb} model yields a lower C-statistic than the power law, although the difference decreases when the fit is restricted to 0.5--6.0~keV. The inferred intrinsic luminosity depends on the spectral model and fitted absorption, but X32 remains in the ultraluminous regime in all cases. These properties are compatible with super-Eddington accretion, although the available spectra do not distinguish between a neutron-star and black-hole accretor.

The epoch-resolved XLFs have cumulative slopes of $\beta\simeq0.66$--$1.00$ above $10^{38}~{\rm erg~s^{-1}}$, broadly consistent with the shape expected for an HMXB population. For the adopted SFR of $1.4~M_\odot\,{\rm yr}^{-1}$, the SFR-scaled HMXB XLF
\citep{2003MNRAS.339..793G,2012MNRAS.419.2095M} predicts approximately
1.1 sources above $10^{39}~{\rm erg~s^{-1}}$, whereas the observed number varies from zero to four between epochs. Considering the intrinsic scatter of the HMXB XLF and the small-number statistics, the observed ULX population does not require an enhanced production of HMXBs in NGC~1300. Instead, the changing number of sources above $10^{39}~{\rm erg~s^{-1}}$ shows that long-term variability strongly modifies the bright end of the observed XLF. Evidence for a broken power law is limited to a small number of epochs and cannot be interpreted as an intrinsic population feature without a completeness correction. The expected contribution of background AGNs is substantial at the faint end of the observed X-ray source population.

Future work will focus on identifying the donor-star counterparts of the ULX population using the \textit{Hubble Space Telescope} (HST) and the \textit{James Webb Space Telescope} (JWST), and on constraining their stellar properties. The sources that can be reliably resolved with the \textit{XMM-Newton} and \textit{Swift}/XRT are also promising targets for future deep X-ray monitoring, which will extend the temporal baseline and provide tighter constraints on their long-term spectral and variability properties.

\begin{acknowledgements}
 This work is partially supported by the Bundesministerium f\"ur Wirtschaft und Energie through the Deutsches Zentrum f\"ur Luft- und Raumfahrt e.V. (DLR) under the grant 50 OR 2517. We thank the anonymous referee for the careful review and constructive comments. We also thank A. Akyuz for helpful suggestions and guidance that improved the manuscript.
\end{acknowledgements}

\bibpunct{(}{)}{;}{a}{}{,}
\bibliographystyle{aa63273-26}
\bibliography{aa63273-26}

@ARTICLE{2009A&A...497..635C,
       author = {{Cappelluti}, N. and {Brusa}, M. and {Hasinger}, G. and {Comastri}, A. and {Zamorani}, G. and {Finoguenov}, A. and {Gilli}, R. and {Puccetti}, S. and {Miyaji}, T. and {Salvato}, M. and {Vignali}, C. and {Aldcroft}, T. and {B{\"o}hringer}, H. and {Brunner}, H. and {Civano}, F. and {Elvis}, M. and {Fiore}, F. and {Fruscione}, A. and {Griffiths}, R.~E. and {Guzzo}, L. and {Iovino}, A. and {Koekemoer}, A.~M. and {Mainieri}, V. and {Scoville}, N.~Z. and {Shopbell}, P. and {Silverman}, J. and {Urry}, C.~M.},
        title = "{The XMM-Newton wide-field survey in the COSMOS field. The point-like X-ray source catalogue}",
      journal = {\aap},
         year = 2009,
        month = apr,
       volume = {497},
       number = {2},
        pages = {635-648},
          doi = {10.1051/0004-6361/200810794},
archivePrefix = {arXiv},
       eprint = {0901.2347},
 primaryClass = {astro-ph.HE},
       adsurl = {https://ui.adsabs.harvard.edu/abs/2009A&A...497..635C}
}

@ARTICLE{2004MNRAS.349..146G,
       author = {{Gilfanov}, M.},
        title = "{Low-mass X-ray binaries as a stellar mass indicator for the host galaxy}",
      journal = {\mnras},
         year = 2004,
        month = mar,
       volume = {349},
       number = {1},
        pages = {146-168},
          doi = {10.1111/j.1365-2966.2004.07473.x},
archivePrefix = {arXiv},
       eprint = {astro-ph/0309454},
 primaryClass = {astro-ph},
       adsurl = {https://ui.adsabs.harvard.edu/abs/2004MNRAS.349..146G}
}

@ARTICLE{2006ARA&A..44..323F,
       author = {{Fabbiano}, G.},
        title = "{Populations of X-Ray Sources in Galaxies}",
      journal = {\araa},
         year = 2006,
        month = sep,
       volume = {44},
       number = {1},
        pages = {323-366},
          doi = {10.1146/annurev.astro.44.051905.092519},
archivePrefix = {arXiv},
       eprint = {astro-ph/0511481},
 primaryClass = {astro-ph},
       adsurl = {https://ui.adsabs.harvard.edu/abs/2006ARA&A..44..323F}
}

@ARTICLE{2012MNRAS.419.2095M,
       author = {{Mineo}, S. and {Gilfanov}, M. and {Sunyaev}, R.},
        title = "{X-ray emission from star-forming galaxies - I. High-mass X-ray binaries}",
      journal = {\mnras},
         year = 2012,
        month = jan,
       volume = {419},
       number = {3},
        pages = {2095-2115},
          doi = {10.1111/j.1365-2966.2011.19862.x},
archivePrefix = {arXiv},
       eprint = {1105.4610},
 primaryClass = {astro-ph.HE},
       adsurl = {https://ui.adsabs.harvard.edu/abs/2012MNRAS.419.2095M}
}

@BOOK{1981rsac.book.....S,
       author = {{Sandage}, Allan and {Tammann}, G.~A.},
        title = "{A Revised Shapley-Ames Catalog of Bright Galaxies}",
         year = 1981,
       adsurl = {https://ui.adsabs.harvard.edu/abs/1981rsac.book.....S}
}

@ARTICLE{1989ApJ...337..191E,
       author = {{England}, Martin N.},
        title = "{High-Resolution Neutral Hydrogen Observations: The Barred Spiral Galaxy NGC 1300}",
      journal = {\apj},
         year = 1989,
        month = feb,
       volume = {337},
        pages = {191},
          doi = {10.1086/167097},
       adsurl = {https://ui.adsabs.harvard.edu/abs/1989ApJ...337..191E}
}

@ARTICLE{2020MNRAS.498.4790K,
       author = {{Kovlakas}, K. and {Zezas}, A. and {Andrews}, J.~J. and {Basu-Zych}, A. and {Fragos}, T. and {Hornschemeier}, A. and {Lehmer}, B. and {Ptak}, A.},
        title = "{A census of ultraluminous X-ray sources in the local Universe}",
      journal = {\mnras},
         year = 2020,
        month = nov,
       volume = {498},
       number = {4},
        pages = {4790-4810},
          doi = {10.1093/mnras/staa2481},
archivePrefix = {arXiv},
       eprint = {2008.10572},
 primaryClass = {astro-ph.GA},
       adsurl = {https://ui.adsabs.harvard.edu/abs/2020MNRAS.498.4790K}
}

@ARTICLE{2013ApJ...769...92P,
       author = {{Prestwich}, A.~H. and {Tsantaki}, Maria and {Zezas}, A. and {Jackson}, F. and {Roberts}, T.~P. and {Foltz}, R. and {Linden}, T. and {Kalogera}, V.},
        title = "{Ultra-luminous X-Ray Sources in the Most Metal Poor Galaxies}",
      journal = {\apj},
         year = 2013,
        month = jun,
       volume = {769},
       number = {2},
          eid = {92},
        pages = {92},
          doi = {10.1088/0004-637X/769/2/92},
archivePrefix = {arXiv},
       eprint = {1302.6203},
 primaryClass = {astro-ph.HE},
       adsurl = {https://ui.adsabs.harvard.edu/abs/2013ApJ...769...92P}
}

@ARTICLE{2024ApJ...977..189L,
       author = {{Lehmer}, Bret D. and {Monson}, Erik B. and {Eufrasio}, Rafael T. and {Amiri}, Amirnezam and {Doore}, Keith and {Basu-Zych}, Antara and {Garofali}, Kristen and {Oskinova}, Lidia and {Andrews}, Jeff J. and {Antoniou}, Vallia and {Geda}, Robel and {Greene}, Jenny E. and {Kovlakas}, Konstantinos and {Lazzarini}, Margaret and {Richardson}, Chris T.},
        title = "{An Empirical Framework Characterizing the Metallicity and Star-formation History Dependence of X-Ray Binary Population Formation and Emission in Galaxies}",
      journal = {\apj},
         year = 2024,
        month = dec,
       volume = {977},
       number = {2},
          eid = {189},
        pages = {189},
          doi = {10.3847/1538-4357/ad8de7},
archivePrefix = {arXiv},
       eprint = {2410.19901},
 primaryClass = {astro-ph.GA},
       adsurl = {https://ui.adsabs.harvard.edu/abs/2024ApJ...977..189L}
}

@ARTICLE{2016MNRAS.459..528A,
       author = {{Antoniou}, V. and {Zezas}, A.},
        title = "{Star formation history and X-ray binary populations: the case of the Large Magellanic Cloud}",
      journal = {\mnras},
         year = 2016,
        month = jun,
       volume = {459},
       number = {1},
        pages = {528-553},
          doi = {10.1093/mnras/stw167},
archivePrefix = {arXiv},
       eprint = {1603.08011},
 primaryClass = {astro-ph.HE},
       adsurl = {https://ui.adsabs.harvard.edu/abs/2016MNRAS.459..528A}
}

@ARTICLE{2020A&A...644A.116R,
       author = {{Rosado-Belza}, D. and {Falc{\'o}n-Barroso}, J. and {Knapen}, J.~H. and {Bittner}, A. and {Gadotti}, D.~A. and {Neumann}, J. and {de Lorenzo-C{\'a}ceres}, A. and {M{\'e}ndez-Abreu}, J. and {Querejeta}, M. and {Mart{\'\i}n-Navarro}, I. and {S{\'a}nchez-Bl{\'a}zquez}, P. and {Coelho}, P.~R.~T. and {Martig}, M. and {van de Ven}, G. and {Kim}, T.},
        title = "{The kinematics of young and old stellar populations in nuclear rings of MUSE TIMER galaxies}",
      journal = {\aap},
         year = 2020,
        month = dec,
       volume = {644},
          eid = {A116},
        pages = {A116},
          doi = {10.1051/0004-6361/202039530},
archivePrefix = {arXiv},
       eprint = {2010.11815},
 primaryClass = {astro-ph.GA},
       adsurl = {https://ui.adsabs.harvard.edu/abs/2020A&A...644A.116R}
}

@ARTICLE{2009ApJ...696.1712F,
       author = {{Feng}, Hua and {Kaaret}, Philip},
        title = "{Spectral States and Evolution of Ultraluminous X-Ray Sources}",
      journal = {\apj},
         year = 2009,
        month = may,
       volume = {696},
       number = {2},
        pages = {1712-1726},
          doi = {10.1088/0004-637X/696/2/1712},
archivePrefix = {arXiv},
       eprint = {0902.2607},
 primaryClass = {astro-ph.HE},
       adsurl = {https://ui.adsabs.harvard.edu/abs/2009ApJ...696.1712F}
}

@ARTICLE{2018MNRAS.477L..90P,
       author = {{Pintore}, F. and {Belfiore}, A. and {Novara}, G. and {Salvaterra}, R. and {Marelli}, M. and {De Luca}, A. and {Rigoselli}, M. and {Israel}, G. and {Rodriguez}, G. and {Mereghetti}, S. and {Wolter}, A. and {Walton}, D.~J. and {Fuerst}, F. and {Ambrosi}, E. and {Zampieri}, L. and {Tiengo}, A. and {Salvaggio}, C.},
        title = "{A new ultraluminous X-ray source in the galaxy NGC 5907}",
      journal = {\mnras},
         year = 2018,
        month = jun,
       volume = {477},
       number = {1},
        pages = {L90-L95},
          doi = {10.1093/mnrasl/sly048},
archivePrefix = {arXiv},
       eprint = {1806.03633},
 primaryClass = {astro-ph.HE},
       adsurl = {https://ui.adsabs.harvard.edu/abs/2018MNRAS.477L..90P}
}

@ARTICLE{2006ApJ...642..171L,
       author = {{Liu}, Ji-Feng and {Bregman}, Joel N. and {Irwin}, Jimmy},
        title = "{Ultraluminous X-Ray Sources in Nearby Galaxies from ROSAT HRI Observations. II. Statistical Properties}",
      journal = {\apj},
         year = 2006,
        month = may,
       volume = {642},
       number = {1},
        pages = {171-187},
          doi = {10.1086/500787},
archivePrefix = {arXiv},
       eprint = {astro-ph/0501312},
 primaryClass = {astro-ph},
       adsurl = {https://ui.adsabs.harvard.edu/abs/2006ApJ...642..171L}
}

@ARTICLE{2017ApJ...836..113P,
       author = {{Pintore}, F. and {Zampieri}, L. and {Stella}, L. and {Wolter}, A. and {Mereghetti}, S. and {Israel}, G.~L.},
        title = "{Pulsator-like Spectra from Ultraluminous X-Ray Sources and the Search for More Ultraluminous Pulsars}",
      journal = {\apj},
         year = 2017,
        month = feb,
       volume = {836},
       number = {1},
          eid = {113},
        pages = {113},
          doi = {10.3847/1538-4357/836/1/113},
archivePrefix = {arXiv},
       eprint = {1701.03595},
 primaryClass = {astro-ph.HE},
       adsurl = {https://ui.adsabs.harvard.edu/abs/2017ApJ...836..113P}
}

@ARTICLE{2022MNRAS.515.4669R,
       author = {{Robba}, A. and {Pinto}, C. and {Pintore}, F. and {Rodriguez}, G. and {Ambrosi}, E. and {Barra}, F. and {Cusumano}, G. and {D'A{\`\i}}, A. and {Del Santo}, M. and {Kosec}, P. and {Marino}, A. and {Middleton}, M. and {Roberts}, T. and {Salvaggio}, C. and {Soria}, R. and {Wolter}, A. and {Walton}, D.},
        title = "{A transient ultraluminous X-ray source in NGC 55}",
      journal = {\mnras},
         year = 2022,
        month = oct,
       volume = {515},
       number = {4},
        pages = {4669-4674},
          doi = {10.1093/mnras/stac2018},
archivePrefix = {arXiv},
       eprint = {2207.09447},
 primaryClass = {astro-ph.HE},
       adsurl = {https://ui.adsabs.harvard.edu/abs/2022MNRAS.515.4669R}
}

@ARTICLE{2019ApJS..243....3L,
       author = {{Lehmer}, Bret D. and {Eufrasio}, Rafael T. and {Tzanavaris}, Panayiotis and {Basu-Zych}, Antara and {Fragos}, Tassos and {Prestwich}, Andrea and {Yukita}, Mihoko and {Zezas}, Andreas and {Hornschemeier}, Ann E. and {Ptak}, Andrew},
        title = "{X-Ray Binary Luminosity Function Scaling Relations for Local Galaxies Based on Subgalactic Modeling}",
      journal = {\apjs},
         year = 2019,
        month = jul,
       volume = {243},
       number = {1},
          eid = {3},
        pages = {3},
          doi = {10.3847/1538-4365/ab22a8},
archivePrefix = {arXiv},
       eprint = {1905.05197},
 primaryClass = {astro-ph.GA},
       adsurl = {https://ui.adsabs.harvard.edu/abs/2019ApJS..243....3L}
}

@ARTICLE{2018PASJ...70...37M,
       author = {{Maeda}, Fumiya and {Ohta}, Kouji and {Fujimoto}, Yusuke and {Habe}, Asao and {Baba}, Junichi},
        title = "{Large velocity dispersion of molecular gas in bars of strongly barred galaxies NGC 1300 and NGC 5383}",
      journal = {\pasj},
         year = 2018,
        month = jun,
       volume = {70},
       number = {3},
          eid = {37},
        pages = {37},
          doi = {10.1093/pasj/psy028},
archivePrefix = {arXiv},
       eprint = {1802.09532},
 primaryClass = {astro-ph.GA},
       adsurl = {https://ui.adsabs.harvard.edu/abs/2018PASJ...70...37M}
}

@ARTICLE{2020MNRAS.494.2131F,
       author = {{Fujimoto}, Yusuke and {Maeda}, Fumiya and {Habe}, Asao and {Ohta}, Kouji},
        title = "{Fast cloud-cloud collisions in a strongly barred galaxy: suppression of massive star formation}",
      journal = {\mnras},
         year = 2020,
        month = may,
       volume = {494},
       number = {2},
        pages = {2131-2146},
          doi = {10.1093/mnras/staa840},
archivePrefix = {arXiv},
       eprint = {2003.12074},
 primaryClass = {astro-ph.GA},
       adsurl = {https://ui.adsabs.harvard.edu/abs/2020MNRAS.494.2131F}
}

@ARTICLE{2020MNRAS.495.3840M,
       author = {{Maeda}, Fumiya and {Ohta}, Kouji and {Fujimoto}, Yusuke and {Habe}, Asao and {Ushio}, Kaito},
        title = "{A large amount of diffuse molecular gases in the bar of the strongly barred galaxy NGC 1300: cause of the low star formation efficiency}",
      journal = {\mnras},
         year = 2020,
        month = jul,
       volume = {495},
       number = {4},
        pages = {3840-3858},
          doi = {10.1093/mnras/staa1296},
archivePrefix = {arXiv},
       eprint = {2005.03019},
 primaryClass = {astro-ph.GA},
       adsurl = {https://ui.adsabs.harvard.edu/abs/2020MNRAS.495.3840M}
}

@ARTICLE{2021ApJS..257...43L,
       author = {{Leroy}, Adam K. and {Schinnerer}, Eva and {Hughes}, Annie and {Rosolowsky}, Erik and {Pety}, J{\'e}r{\^o}me and {Schruba}, Andreas and {Usero}, Antonio and {Blanc}, Guillermo A. and {Chevance}, M{\'e}lanie and {Emsellem}, Eric and {Faesi}, Christopher M. and {Herrera}, Cinthya N. and {Liu}, Daizhong and {Meidt}, Sharon E. and {Querejeta}, Miguel and {Saito}, Toshiki and {Sandstrom}, Karin M. and {Sun}, Jiayi and {Williams}, Thomas G. and {Anand}, Gagandeep S. and {Barnes}, Ashley T. and {Behrens}, Erica A. and {Belfiore}, Francesco and {Benincasa}, Samantha M. and {Be{\v{s}}li{\'c}}, Ivana and {Bigiel}, Frank and {Bolatto}, Alberto D. and {den Brok}, Jakob S. and {Cao}, Yixian and {Chandar}, Rupali and {Chastenet}, J{\'e}r{\'e}my and {Chiang}, I-Da and {Congiu}, Enrico and {Dale}, Daniel A. and {Deger}, Sinan and {Eibensteiner}, Cosima and {Egorov}, Oleg V. and {Garc{\'\i}a-Rodr{\'\i}guez}, Axel and {Glover}, Simon C.~O. and {Grasha}, Kathryn and {Henshaw}, Jonathan D. and {Ho}, I.-Ting and {Kepley}, Amanda A. and {Kim}, Jaeyeon and {Klessen}, Ralf S. and {Kreckel}, Kathryn and {Koch}, Eric W. and {Kruijssen}, J.~M. Diederik and {Larson}, Kirsten L. and {Lee}, Janice C. and {Lopez}, Laura A. and {Machado}, Josh and {Mayker}, Ness and {McElroy}, Rebecca and {Murphy}, Eric J. and {Ostriker}, Eve C. and {Pan}, Hsi-An and {Pessa}, Ismael and {Puschnig}, Johannes and {Razza}, Alessandro and {S{\'a}nchez-Bl{\'a}zquez}, Patricia and {Santoro}, Francesco and {Sardone}, Amy and {Scheuermann}, Fabian and {Sliwa}, Kazimierz and {Sormani}, Mattia C. and {Stuber}, Sophia K. and {Thilker}, David A. and {Turner}, Jordan A. and {Utomo}, Dyas and {Watkins}, Elizabeth J. and {Whitmore}, Bradley},
        title = "{PHANGS-ALMA: Arcsecond CO(2-1) Imaging of Nearby Star-forming Galaxies}",
      journal = {\apjs},
         year = 2021,
        month = dec,
       volume = {257},
       number = {2},
          eid = {43},
        pages = {43},
          doi = {10.3847/1538-4365/ac17f3},
archivePrefix = {arXiv},
       eprint = {2104.07739},
 primaryClass = {astro-ph.GA},
       adsurl = {https://ui.adsabs.harvard.edu/abs/2021ApJS..257...43L}
}

@ARTICLE{2020MNRAS.493.5045M,
       author = {{Maeda}, Fumiya and {Ohta}, Kouji and {Fujimoto}, Yusuke and {Habe}, Asao},
        title = "{Properties of giant molecular clouds in the strongly barred galaxy NGC 1300}",
      journal = {\mnras},
         year = 2020,
        month = apr,
       volume = {493},
       number = {4},
        pages = {5045-5061},
          doi = {10.1093/mnras/staa556},
archivePrefix = {arXiv},
       eprint = {2002.08977},
 primaryClass = {astro-ph.GA},
       adsurl = {https://ui.adsabs.harvard.edu/abs/2020MNRAS.493.5045M}
}

@ARTICLE{2012ApJ...749..129G,
       author = {{G{\"u}ltekin}, Kayhan and {Cackett}, Edward M. and {Miller}, Jon M. and {Di Matteo}, Tiziana and {Markoff}, Sera and {Richstone}, Douglas O.},
        title = "{A Chandra Survey of Supermassive Black Holes with Dynamical Mass Measurements}",
      journal = {\apj},
         year = 2012,
        month = apr,
       volume = {749},
       number = {2},
          eid = {129},
        pages = {129},
          doi = {10.1088/0004-637X/749/2/129},
archivePrefix = {arXiv},
       eprint = {1202.3147},
 primaryClass = {astro-ph.HE},
       adsurl = {https://ui.adsabs.harvard.edu/abs/2012ApJ...749..129G}
}

@ARTICLE{2003ApJ...595..719P,
       author = {{Prestwich}, A.~H. and {Irwin}, J.~A. and {Kilgard}, R.~E. and {Krauss}, M.~I. and {Zezas}, A. and {Primini}, F. and {Kaaret}, P. and {Boroson}, B.},
        title = "{Classifying X-Ray Sources in External Galaxies from X-Ray Colors}",
      journal = {\apj},
         year = 2003,
        month = oct,
       volume = {595},
       number = {2},
        pages = {719-726},
          doi = {10.1086/377366},
archivePrefix = {arXiv},
       eprint = {astro-ph/0206127},
 primaryClass = {astro-ph},
       adsurl = {https://ui.adsabs.harvard.edu/abs/2003ApJ...595..719P}
}

@ARTICLE{2020A&A...643A.171H,
       author = {{Hameury}, J. -M. and {Lasota}, J. -P.},
        title = "{Models of ultraluminous X-ray transient sources}",
      journal = {\aap},
         year = 2020,
        month = nov,
       volume = {643},
          eid = {A171},
        pages = {A171},
          doi = {10.1051/0004-6361/202038857},
archivePrefix = {arXiv},
       eprint = {2010.00365},
 primaryClass = {astro-ph.HE},
       adsurl = {https://ui.adsabs.harvard.edu/abs/2020A&A...643A.171H}
}

@ARTICLE{2023ApJ...951...51B,
       author = {{Brightman}, Murray and {Hameury}, Jean-Marie and {Lasota}, Jean-Pierre and {Baldi}, Ranieri D. and {Bruni}, Gabriele and {Cann}, Jenna M. and {Earnshaw}, Hannah and {F{\"u}rst}, Felix and {Heida}, Marianne and {Jaodand}, Amruta and {Lazzarini}, Margaret and {Middleton}, Matthew J. and {Walton}, Dominic J. and {Weaver}, Kimberly A.},
        title = "{A New Sample of Transient Ultraluminous X-Ray Sources Serendipitously Discovered by Swift/XRT}",
      journal = {\apj},
         year = 2023,
        month = jul,
       volume = {951},
       number = {1},
          eid = {51},
        pages = {51},
          doi = {10.3847/1538-4357/acd18a},
archivePrefix = {arXiv},
       eprint = {2305.01693},
 primaryClass = {astro-ph.HE},
       adsurl = {https://ui.adsabs.harvard.edu/abs/2023ApJ...951...51B}
}

@ARTICLE{2020ApJ...891..153E,
       author = {{Earnshaw}, Hannah P. and {Heida}, Marianne and {Brightman}, Murray and {F{\"u}rst}, Felix and {Harrison}, Fiona A. and {Jaodand}, Amruta and {Middleton}, Matthew J. and {Roberts}, Timothy P. and {Sathyaprakash}, Rajath and {Stern}, Daniel and {Walton}, Dominic J.},
        title = "{The (Re)appearance of NGC 925 ULX-3, a New Transient ULX}",
      journal = {\apj},
         year = 2020,
        month = mar,
       volume = {891},
       number = {2},
          eid = {153},
        pages = {153},
          doi = {10.3847/1538-4357/ab77b8},
archivePrefix = {arXiv},
       eprint = {2001.00642},
 primaryClass = {astro-ph.HE},
       adsurl = {https://ui.adsabs.harvard.edu/abs/2020ApJ...891..153E}
}

@ARTICLE{2019MNRAS.483.3566V,
       author = {{van Haaften}, Lennart M. and {Maccarone}, Thomas J. and {Rhode}, Katherine L. and {Kundu}, Arunav and {Zepf}, Stephen E.},
        title = "{Discovery of a transient ultraluminous X-ray source in the elliptical galaxy M86}",
      journal = {\mnras},
         year = 2019,
        month = mar,
       volume = {483},
       number = {3},
        pages = {3566-3573},
          doi = {10.1093/mnras/sty3221},
archivePrefix = {arXiv},
       eprint = {1811.10595},
 primaryClass = {astro-ph.HE},
       adsurl = {https://ui.adsabs.harvard.edu/abs/2019MNRAS.483.3566V}
}

@ARTICLE{2020MNRAS.494.3611K,
       author = {{King}, Andrew and {Lasota}, Jean-Pierre},
        title = "{Pulsing and non-pulsing ULXs: the iceberg emerges}",
      journal = {\mnras},
         year = 2020,
        month = may,
       volume = {494},
       number = {3},
        pages = {3611-3615},
          doi = {10.1093/mnras/staa930},
archivePrefix = {arXiv},
       eprint = {2003.14019},
 primaryClass = {astro-ph.HE},
       adsurl = {https://ui.adsabs.harvard.edu/abs/2020MNRAS.494.3611K}
}

@ARTICLE{2013MNRAS.435.1758S,
       author = {{Sutton}, Andrew D. and {Roberts}, Timothy P. and {Middleton}, Matthew J.},
        title = "{The ultraluminous state revisited: fractional variability and spectral shape as diagnostics of super-Eddington accretion}",
      journal = {\mnras},
         year = 2013,
        month = oct,
       volume = {435},
       number = {2},
        pages = {1758-1775},
          doi = {10.1093/mnras/stt1419},
archivePrefix = {arXiv},
       eprint = {1307.8044},
 primaryClass = {astro-ph.HE},
       adsurl = {https://ui.adsabs.harvard.edu/abs/2013MNRAS.435.1758S}
}

@ARTICLE{2003MNRAS.339..793G,
       author = {{Grimm}, H. -J. and {Gilfanov}, M. and {Sunyaev}, R.},
        title = "{High-mass X-ray binaries as a star formation rate indicator in distant galaxies}",
      journal = {\mnras},
         year = 2003,
        month = mar,
       volume = {339},
       number = {3},
        pages = {793-809},
          doi = {10.1046/j.1365-8711.2003.06224.x},
archivePrefix = {arXiv},
       eprint = {astro-ph/0205371},
 primaryClass = {astro-ph},
       adsurl = {https://ui.adsabs.harvard.edu/abs/2003MNRAS.339..793G}
}

@ARTICLE{2018ApJ...856..128W,
       author = {{Walton}, D.~J. and {F{\"u}rst}, F. and {Heida}, M. and {Harrison}, F.~A. and {Barret}, D. and {Stern}, D. and {Bachetti}, M. and {Brightman}, M. and {Fabian}, A.~C. and {Middleton}, M.~J.},
        title = "{Evidence for Pulsar-like Emission Components in the Broadband ULX Sample}",
      journal = {\apj},
         year = 2018,
        month = apr,
       volume = {856},
       number = {2},
          eid = {128},
        pages = {128},
          doi = {10.3847/1538-4357/aab610},
archivePrefix = {arXiv},
       eprint = {1803.04424},
 primaryClass = {astro-ph.HE},
       adsurl = {https://ui.adsabs.harvard.edu/abs/2018ApJ...856..128W}
}

@ARTICLE{2023MNRAS.526.5765A,
       author = {{Allak}, Sinan},
        title = "{New transient ULX candidate in NGC 4254: evidence of circumbinary disc?}",
      journal = {\mnras},
         year = 2023,
        month = dec,
       volume = {526},
       number = {4},
        pages = {5765-5776},
          doi = {10.1093/mnras/stad3175},
archivePrefix = {arXiv},
       eprint = {2310.09421},
       adsurl = {https://ui.adsabs.harvard.edu/abs/2023MNRAS.526.5765A}
}

@ARTICLE{2021AstBu..76....6F,
       author = {{Fabrika}, S.~N. and {Atapin}, K.~E. and {Vinokurov}, A.~S. and {Sholukhova}, O.~N.},
        title = "{Ultraluminous X-Ray Sources}",
      journal = {Astrophysical Bulletin},
         year = 2021,
        month = jan,
       volume = {76},
       number = {1},
        pages = {6-38},
          doi = {10.1134/S1990341321010077},
archivePrefix = {arXiv},
       eprint = {2105.10537},
 primaryClass = {astro-ph.GA},
       adsurl = {https://ui.adsabs.harvard.edu/abs/2021AstBu..76....6F}
}

@ARTICLE{2014Natur.514..202B,
	author = {{Bachetti}, M. and {Harrison}, F.~A. and {Walton}, D.~J. and 
	{Grefenstette}, B.~W. and {Chakrabarty}, D. and {F{\"u}rst}, F. and 
	{Barret}, D. and {Beloborodov}, A. and {Boggs}, S.~E. and {Christensen}, F.~E. and 
	{Craig}, W.~W. and {Fabian}, A.~C. and {Hailey}, C.~J. and {Hornschemeier}, A. and 
	{Kaspi}, V. and {Kulkarni}, S.~R. and {Maccarone}, T. and {Miller}, J.~M. and 
	{Rana}, V. and {Stern}, D. and {Tendulkar}, S.~P. and {Tomsick}, J. and 
	{Webb}, N.~A. and {Zhang}, W.~W.},
	title = "{An ultraluminous X-ray source powered by an accreting neutron star}",
	journal = {\nat},
	archivePrefix = "arXiv",
	eprint = {1410.3590},
	primaryClass = "astro-ph.HE",
	year = 2014,
	month = oct,
	volume = 514,
	pages = {202-204},
	doi = {10.1038/nature13791},
	adsurl = {http://adsabs.harvard.edu/abs/2014Natur.514..202B}
}

@ARTICLE{2007A&A...469..379E,
       author = {{Evans}, P.~A. and {Beardmore}, A.~P. and {Page}, K.~L. and {Tyler}, L.~G. and {Osborne}, J.~P. and {Goad}, M.~R. and {O'Brien}, P.~T. and {Vetere}, L. and {Racusin}, J. and {Morris}, D. and {Burrows}, D.~N. and {Capalbi}, M. and {Perri}, M. and {Gehrels}, N. and {Romano}, P.},
        title = "{An online repository of Swift/XRT light curves of {\ensuremath{\gamma}}-ray bursts}",
      journal = {\aap},
         year = 2007,
        month = jul,
       volume = {469},
       number = {1},
        pages = {379-385},
          doi = {10.1051/0004-6361:20077530},
archivePrefix = {arXiv},
       eprint = {0704.0128},
 primaryClass = {astro-ph},
       adsurl = {https://ui.adsabs.harvard.edu/abs/2007A&A...469..379E}
}

@ARTICLE{2009MNRAS.397.1836G,
       author = {{Gladstone}, Jeanette C. and {Roberts}, Timothy P. and {Done}, Chris},
        title = "{The ultraluminous state}",
      journal = {\mnras},
         year = 2009,
        month = aug,
       volume = {397},
       number = {4},
        pages = {1836-1851},
          doi = {10.1111/j.1365-2966.2009.15123.x},
archivePrefix = {arXiv},
       eprint = {0905.4076},
 primaryClass = {astro-ph.CO},
       adsurl = {https://ui.adsabs.harvard.edu/abs/2009MNRAS.397.1836G}
}

@ARTICLE{2017ARA&A..55..303K,
	author = {{Kaaret}, P. and {Feng}, H. and {Roberts}, T.~P.},
	title = "{Ultraluminous X-Ray Sources}",
	journal = {\araa},
	archivePrefix = "arXiv",
	eprint = {1703.10728},
	primaryClass = "astro-ph.HE",
	year = 2017,
	month = aug,
	volume = 55,
	pages = {303-341},
	doi = {10.1146/annurev-astro-091916-055259},
	adsurl = {http://adsabs.harvard.edu/abs/2017ARA%26A..55..303K}
}

@INPROCEEDINGS{2006SPIE.6270E..1VF,
       author = {{Fruscione}, Antonella and {McDowell}, Jonathan C. and {Allen}, Glenn E. and {Brickhouse}, Nancy S. and {Burke}, Douglas J. and {Davis}, John E. and {Durham}, Nick and {Elvis}, Martin and {Galle}, Elizabeth C. and {Harris}, Daniel E. and {Huenemoerder}, David P. and {Houck}, John C. and {Ishibashi}, Bish and {Karovska}, Margarita and {Nicastro}, Fabrizio and {Noble}, Michael S. and {Nowak}, Michael A. and {Primini}, Frank A. and {Siemiginowska}, Aneta and {Smith}, Randall K. and {Wise}, Michael},
        title = "{CIAO: Chandra's data analysis system}",
    booktitle = {Observatory Operations: Strategies, Processes, and Systems},
         year = 2006,
       editor = {{Silva}, David R. and {Doxsey}, Rodger E.},
       series = {Society of Photo-Optical Instrumentation Engineers (SPIE) Conference Series},
       volume = {6270},
        month = jun,
          eid = {62701V},
        pages = {62701V},
          doi = {10.1117/12.671760},
       adsurl = {https://ui.adsabs.harvard.edu/abs/2006SPIE.6270E..1VF}
}

@ARTICLE{2009MNRAS.397.1177E,
       author = {{Evans}, P.~A. and {Beardmore}, A.~P. and {Page}, K.~L. and {Osborne}, J.~P. and {O'Brien}, P.~T. and {Willingale}, R. and {Starling}, R.~L.~C. and {Burrows}, D.~N. and {Godet}, O. and {Vetere}, L. and {Racusin}, J. and {Goad}, M.~R. and {Wiersema}, K. and {Angelini}, L. and {Capalbi}, M. and {Chincarini}, G. and {Gehrels}, N. and {Kennea}, J.~A. and {Margutti}, R. and {Morris}, D.~C. and {Mountford}, C.~J. and {Pagani}, C. and {Perri}, M. and {Romano}, P. and {Tanvir}, N.},
        title = "{Methods and results of an automatic analysis of a complete sample of Swift-XRT observations of GRBs}",
      journal = {\mnras},
         year = 2009,
        month = aug,
       volume = {397},
       number = {3},
        pages = {1177-1201},
          doi = {10.1111/j.1365-2966.2009.14913.x},
archivePrefix = {arXiv},
       eprint = {0812.3662},
 primaryClass = {astro-ph},
       adsurl = {https://ui.adsabs.harvard.edu/abs/2009MNRAS.397.1177E}
}

@ARTICLE{2011ApJ...734...33J,
       author = {{Jenkins}, L.~P. and {Brandt}, W.~N. and {Colbert}, E.~J.~M. and
         {Koribalski}, B. and {Kuntz}, K.~D. and {Levan}, A.~J. and {Ojha}, R. and
         {Roberts}, T.~P. and {Ward}, M.~J. and {Zezas}, A.},
        title = "{Investigating the Nuclear Activity of Barred Spiral Galaxies: The Case of NGC 1672}",
      journal = {\apj},
         year = 2011,
        month = jun,
       volume = {734},
       number = {1},
          eid = {33},
        pages = {33},
          doi = {10.1088/0004-637X/734/1/33},
archivePrefix = {arXiv},
       eprint = {1104.1649},
 primaryClass = {astro-ph.CO},
       adsurl = {https://ui.adsabs.harvard.edu/abs/2011ApJ...734...33J}
}

@ARTICLE{2021MNRAS.501.1002W,
       author = {{Walton}, D.~J. and {Heida}, M. and {Bachetti}, M. and {F{\"u}rst}, F. and {Brightman}, M. and {Earnshaw}, H. and {Evans}, P.~A. and {Fabian}, A.~C. and {Grefenstette}, B.~W. and {Harrison}, F.~A. and {Israel}, G.~L. and {Lansbury}, G.~B. and {Middleton}, M.~J. and {Pike}, S. and {Rana}, V. and {Roberts}, T.~P. and {Rodriguez Castillo}, G.~A. and {Salvaterra}, R. and {Song}, X. and {Stern}, D.},
        title = "{A new transient ultraluminous X-ray source in NGC 7090}",
      journal = {\mnras},
         year = 2021,
        month = feb,
       volume = {501},
       number = {1},
        pages = {1002-1012},
          doi = {10.1093/mnras/staa3666},
archivePrefix = {arXiv},
       eprint = {2011.08870},
 primaryClass = {astro-ph.HE},
       adsurl = {https://ui.adsabs.harvard.edu/abs/2021MNRAS.501.1002W}
}

@ARTICLE{2020MNRAS.491.1260S,
       author = {{Song}, X. and {Walton}, D.~J. and {Lansbury}, G.~B. and {Evans}, P.~A. and {Fabian}, A.~C. and {Earnshaw}, H. and {Roberts}, T.~P.},
        title = "{The hunt for pulsating ultraluminous X-ray sources}",
      journal = {\mnras},
         year = 2020,
        month = jan,
       volume = {491},
       number = {1},
        pages = {1260-1277},
          doi = {10.1093/mnras/stz3036},
archivePrefix = {arXiv},
       eprint = {1910.11857},
 primaryClass = {astro-ph.HE},
       adsurl = {https://ui.adsabs.harvard.edu/abs/2020MNRAS.491.1260S}
}

\appendix

\section{X-ray observation log}
\label{app:observations}

\begin{table}[H]
\centering
\caption{Log of the \textit{Chandra}/ACIS-S and \textit{Swift}/XRT
observations of NGC~1300 used in this study.}
\label{tab:obs}

\resizebox{\hsize}{!}{
\begin{tabular}{lccc}
\hline\hline
Instrument & ObsID & Exposure (ks) & Start date \\
\hline

\multicolumn{4}{c}{\textit{Chandra}/ACIS-S} \\
\hline
\textit{Chandra}/ACIS-S & 11775 & 29.66 & 2009-10-11 \\
\textit{Chandra}/ACIS-S & 28537 & 8.96 & 2024-12-23 \\
\textit{Chandra}/ACIS-S & 30697 & 9.95 & 2024-12-26 \\
\textit{Chandra}/ACIS-S & 30698 & 8.82 & 2024-12-29 \\
\textit{Chandra}/ACIS-S & 28539 & 25.24 & 2025-01-16 \\
\textit{Chandra}/ACIS-S & 28541 & 16.85 & 2025-01-23 \\
\textit{Chandra}/ACIS-S & 30743 & 12.41 & 2025-01-25 \\
\textit{Chandra}/ACIS-S & 28538 & 9.93 & 2025-06-09 \\
\textit{Chandra}/ACIS-S & 30963 & 19.80 & 2025-06-11 \\
\textit{Chandra}/ACIS-S & 28540 & 19.31 & 2025-06-17 \\
\textit{Chandra}/ACIS-S & 30966 & 14.38 & 2025-06-19 \\
\textit{Chandra}/ACIS-S & 28120 & 53.35 & 2025-08-21 \\
\hline

\multicolumn{4}{c}{\textit{Swift}/XRT} \\
\hline
\textit{Swift}/XRT & 00015435001 & 2.00 & 2022-12-06 \\
\textit{Swift}/XRT & 00015435002 & 2.21 & 2022-12-06 \\
\textit{Swift}/XRT & 00015435003 & 1.42 & 2022-12-06 \\
\textit{Swift}/XRT & 00015435010 & 1.76 & 2022-12-07 \\
\textit{Swift}/XRT & 00015435011 & 1.63 & 2022-12-07 \\
\textit{Swift}/XRT & 00015435012 & 1.63 & 2022-12-08 \\
\textit{Swift}/XRT & 00015435014 & 1.61 & 2022-12-09 \\
\textit{Swift}/XRT & 00015435015 & 1.64 & 2022-12-09 \\
\textit{Swift}/XRT & 00015435016 & 1.77 & 2022-12-10 \\
\textit{Swift}/XRT & 00015435017 & 1.87 & 2022-12-10 \\
\textit{Swift}/XRT & 00015435018 & 1.62 & 2022-12-11 \\
\textit{Swift}/XRT & 00015435019 & 1.63 & 2022-12-11 \\
\textit{Swift}/XRT & 00015435020 & 2.24 & 2022-12-12 \\
\textit{Swift}/XRT & 00015435021 & 1.91 & 2022-12-13 \\
\textit{Swift}/XRT & 00015435022 & 1.13 & 2022-12-14 \\
\textit{Swift}/XRT & 00015435023 & 1.58 & 2022-12-15 \\
\textit{Swift}/XRT & 00015435024 & 0.97 & 2022-12-16 \\
\textit{Swift}/XRT & 00015435025 & 1.43 & 2022-12-17 \\
\textit{Swift}/XRT & 00015435026 & 2.15 & 2022-12-18 \\
\textit{Swift}/XRT & 00015435027 & 1.73 & 2022-12-19 \\
\textit{Swift}/XRT & 00015435028 & 1.94 & 2022-12-20 \\
\textit{Swift}/XRT & 00015435029 & 1.64 & 2022-12-21 \\
\textit{Swift}/XRT & 00015435030 & 1.09 & 2022-12-22 \\
\textit{Swift}/XRT & 00015435031 & 1.67 & 2022-12-23 \\
\textit{Swift}/XRT & 00015435032 & 0.93 & 2022-12-24 \\
\textit{Swift}/XRT & 00015435033 & 1.06 & 2022-12-25 \\
\textit{Swift}/XRT & 00015435034 & 2.11 & 2022-12-26 \\
\textit{Swift}/XRT & 00015435035 & 1.03 & 2022-12-27 \\
\textit{Swift}/XRT & 00015435036 & 2.07 & 2022-12-28 \\
\textit{Swift}/XRT & 00015435037 & 1.58 & 2022-12-31 \\
\textit{Swift}/XRT & 00015435038 & 1.55 & 2023-01-02 \\
\textit{Swift}/XRT & 00015435039 & 1.64 & 2023-01-04 \\
\textit{Swift}/XRT & 00015435040 & 1.68 & 2023-01-06 \\
\textit{Swift}/XRT & 00015435041 & 2.79 & 2023-06-14 \\
\textit{Swift}/XRT & 00015435042 & 4.55 & 2023-07-12 \\
\textit{Swift}/XRT & 00015435043 & 3.28 & 2023-08-23 \\
\textit{Swift}/XRT & 00015435044 & 4.04 & 2023-08-24 \\
\hline
\end{tabular}
}

\tablefoot{Exposure times correspond to the effective exposures of the
individual observations. The observations are listed chronologically
within each instrument.}

\end{table}

\section{X-ray source positions}
\label{app:positions}

\begin{table}[H]
\centering
\caption{J2000 coordinates of the X-ray sources detected within the
$D_{25}$ region of NGC~1300.}
\label{tab:xray_sources}

\small
\begin{tabular}{lcc}
\hline\hline
Source & R.A. (J2000) & Decl. (J2000) \\
\hline
X1 & 03:19:39.46 & $-$19:26:15.70 \\
X2 & 03:19:37.89 & $-$19:26:08.19 \\
X3 & 03:19:32.44 & $-$19:26:06.36 \\
X4 & 03:19:42.30 & $-$19:25:55.15 \\
X5 & 03:19:39.96 & $-$19:25:50.57 \\
X6 & 03:19:42.93 & $-$19:25:11.30 \\
X7 & 03:19:38.56 & $-$19:24:52.71 \\
X8 & 03:19:42.51 & $-$19:24:50.74 \\
X9 & 03:19:45.95 & $-$19:24:49.57 \\
X10 & 03:19:42.43 & $-$19:24:45.04 \\
X11 & 03:19:37.96 & $-$19:24:41.71 \\
X12 & 03:19:41.06 & $-$19:24:40.31 \\
X13 & 03:19:39.38 & $-$19:24:23.34 \\
X14 & 03:19:35.54 & $-$19:24:17.57 \\
X15 & 03:19:38.21 & $-$19:24:13.62 \\
X16 & 03:19:41.38 & $-$19:23:34.42 \\
X17 & 03:19:45.76 & $-$19:22:55.73 \\
X18 & 03:19:47.58 & $-$19:22:35.62 \\
X19 & 03:19:45.59 & $-$19:22:05.43 \\
X20 & 03:19:42.31 & $-$19:27:01.76 \\
X21 & 03:19:35.06 & $-$19:25:10.01 \\
X22 & 03:19:35.09 & $-$19:24:17.68 \\
X23 & 03:19:38.46 & $-$19:23:17.97 \\
X24 & 03:19:38.96 & $-$19:24:55.05 \\
X25 & 03:19:47.57 & $-$19:26:15.42 \\
X26 & 03:19:43.20 & $-$19:24:17.63 \\
X27 & 03:19:39.32 & $-$19:25:46.01 \\
X28 & 03:19:48.37 & $-$19:24:15.65 \\
X29 & 03:19:40.39 & $-$19:24:44.62 \\
X30 & 03:19:38.73 & $-$19:23:39.65 \\
X31 & 03:19:38.35 & $-$19:23:46.22 \\
X32 & 03:19:40.60 & $-$19:24:01.68 \\
X33 & 03:19:41.18 & $-$19:23:37.82 \\
X34 & 03:19:41.88 & $-$19:27:07.20 \\
X35 & 03:19:46.98 & $-$19:24:41.80 \\
X36 & 03:19:39.52 & $-$19:22:41.59 \\
X37 & 03:19:45.03 & $-$19:23:44.04 \\
X38 & 03:19:45.66 & $-$19:24:01.94 \\
X39 & 03:19:47.90 & $-$19:25:11.61 \\
X40 & 03:19:36.88 & $-$19:26:55.46 \\
X41 & 03:19:48.54 & $-$19:24:59.08 \\
X42 & 03:19:48.50 & $-$19:25:20.64 \\
\hline
\end{tabular}
\end{table}


\section{X-ray luminosity distributions of the sources in NGC~1300}

\begin{figure*}
\sidecaption
\centering
\includegraphics[width=12cm]{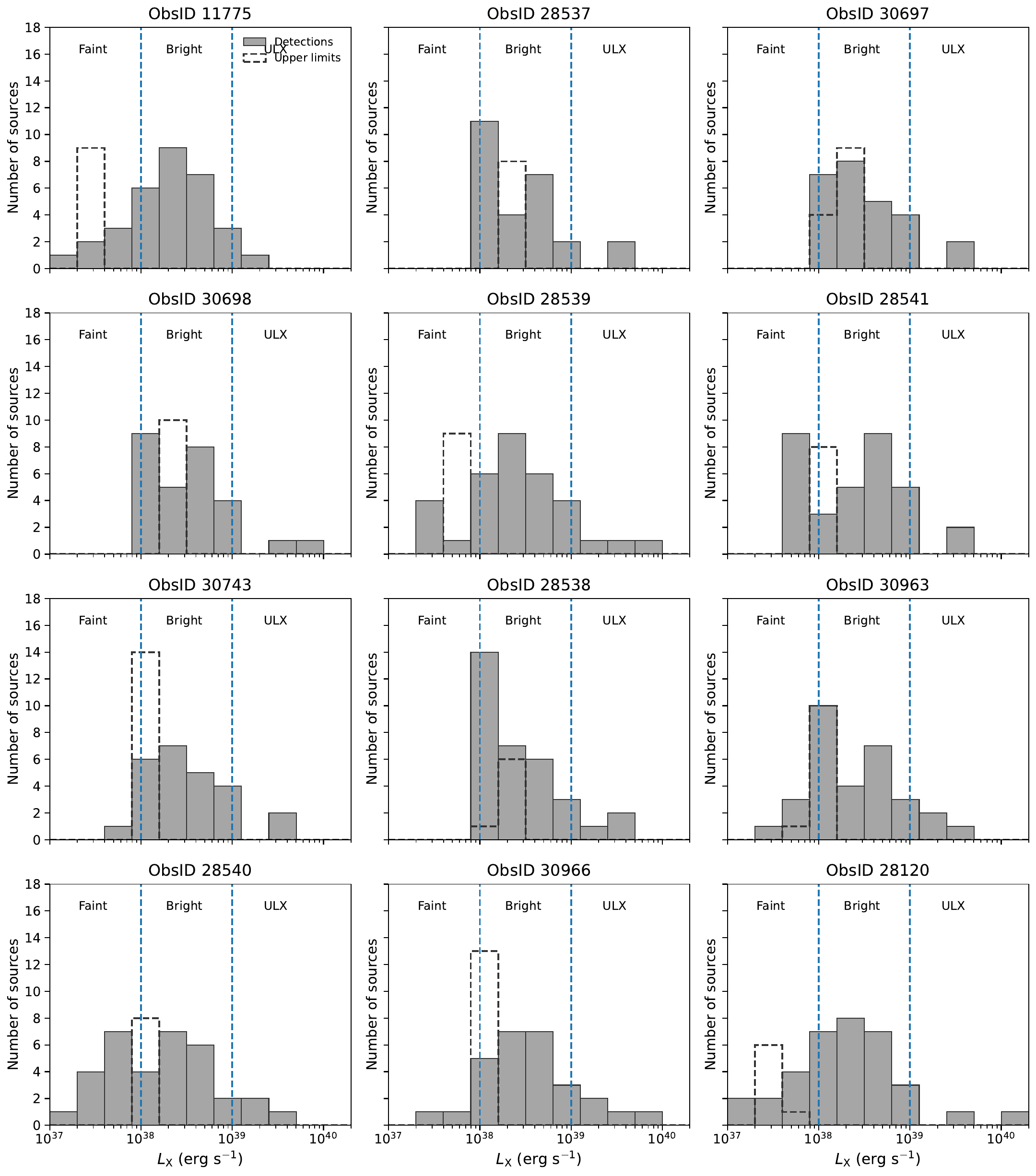}
\caption{X-ray luminosity distributions of the sources in NGC~1300 for the 12 individual \textit{Chandra} observations. The gray filled histograms show the measured 0.3--10~keV luminosities, $L_{\rm X}$, while the black dashed step histograms indicate the corresponding upper-limit distributions for non-detections. Identical logarithmic luminosity bins are used in all panels to facilitate a direct comparison between epochs. The blue dashed vertical lines mark $L_{\rm X}=10^{38}$ and $10^{39}~{\rm erg~s^{-1}}$, separating the adopted faint, bright, and ULX luminosity regimes.}
\label{fig:lx_hist}
\end{figure*}


\section{Long-term X-ray light curves}

\begin{figure*}
\sidecaption
\centering
\includegraphics[width=12cm]{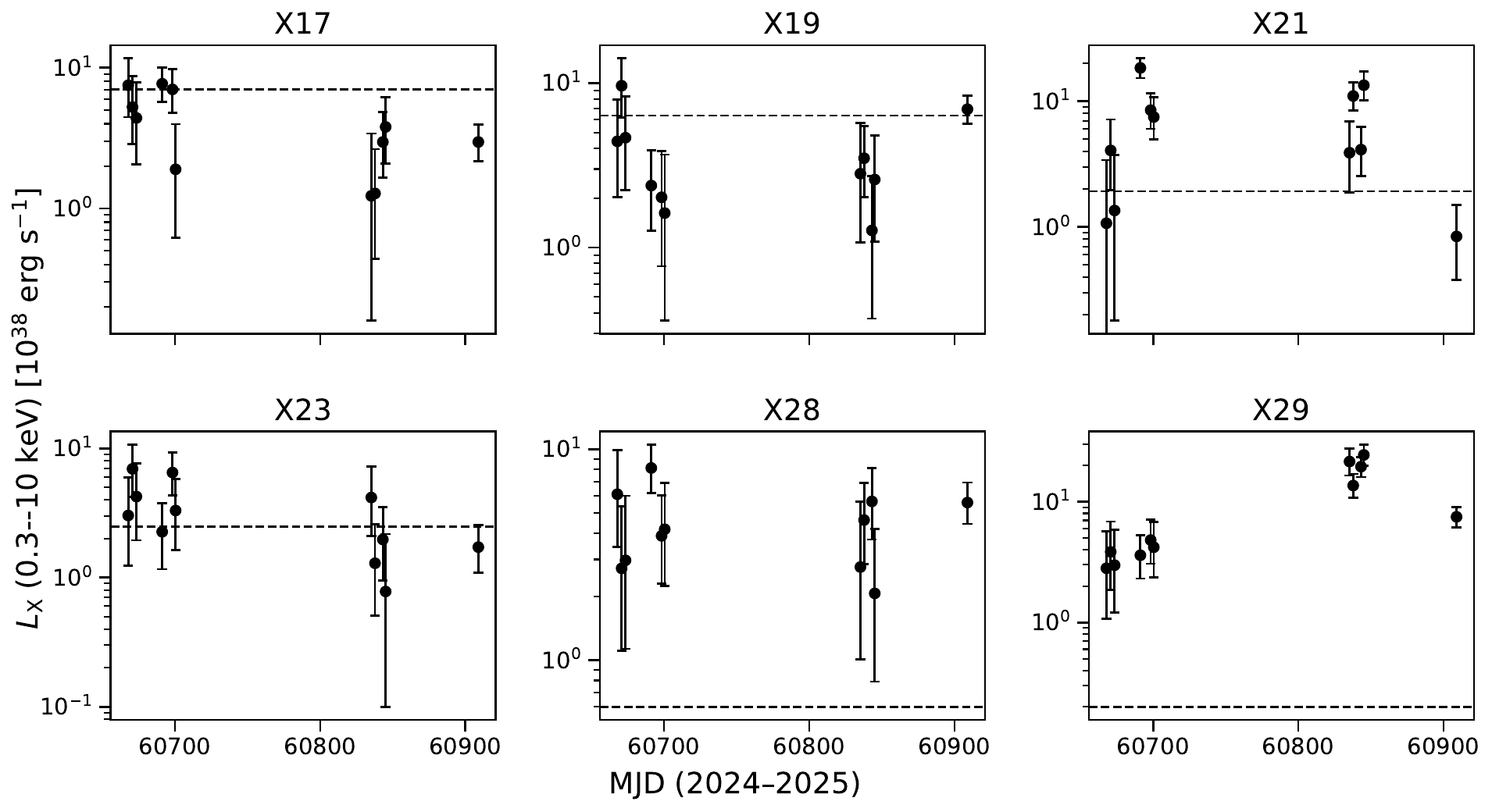}
\caption{Long-term 0.3--10~keV X-ray light curves of six selected sources in NGC~1300 during the 2024--2025 \textit{Chandra} monitoring campaign. From left to right, the upper panels show X17, X19, and X21, while the lower panels show X23, X28, and X29. Luminosities are plotted as a function of MJD, with asymmetric uncertainties shown by the error bars. The horizontal dashed line in each panel indicates the luminosity measured in the archival 2009 \textit{Chandra} observation (ObsID~11775).}
\label{fig:longterm_lc}
\end{figure*}

\section{Swift/XRT image of NGC~1300}

\begin{figure}
\resizebox{\hsize}{!}{\includegraphics{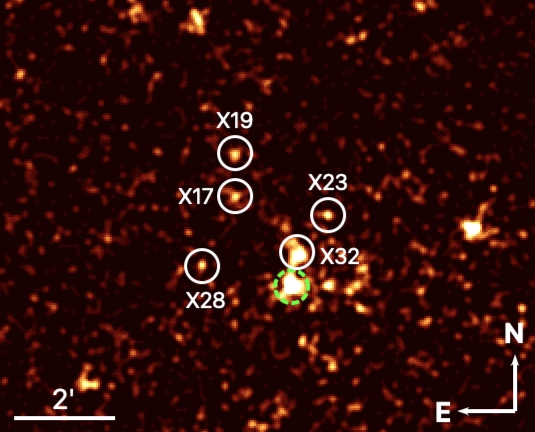}}
\caption{Stacked 0.3--10~keV \textit{Swift}/XRT image of NGC~1300. The white circles, each with a radius of $20\arcsec$, mark the \textit{Chandra}-resolved sources X17, X19, X23, X28, and X32 used in the long-term variability analysis. The green dashed circle marks the center of NGC~1300. North is up and east is to the left.}
\label{fig:xrt}
\end{figure}

\section{X-ray color--color diagram of X32}

\begin{figure}
\resizebox{\hsize}{!}{\includegraphics{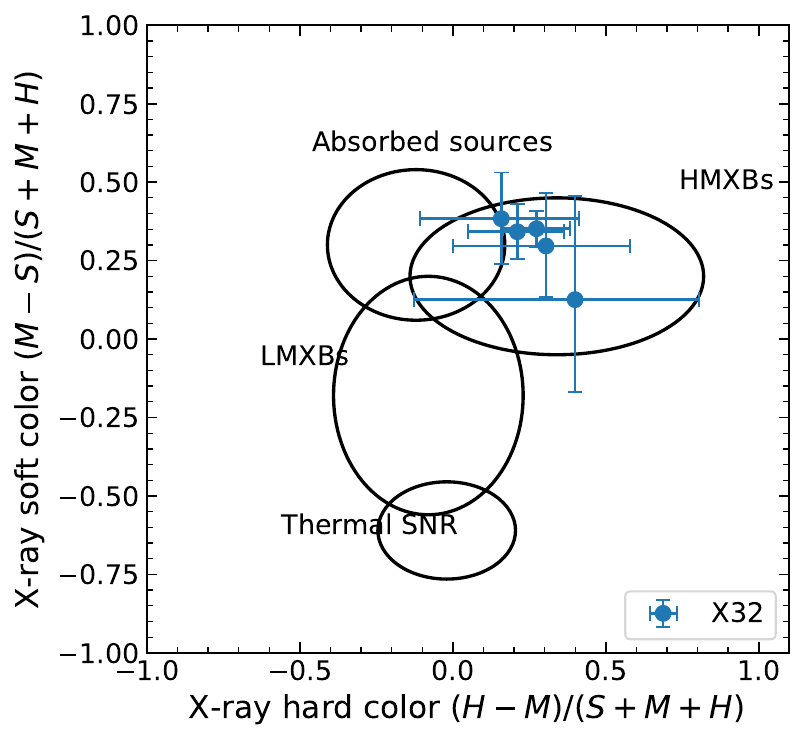}}
\caption{X-ray color--color diagram of X32. The soft and hard colors are defined as $(M-S)/(S+M+H)$ and $(H-M)/(S+M+H)$, respectively, where $S$, $M$, and $H$ correspond to the 0.3--1.0, 1.0--2.0, and 2.0--8.0~keV bands. The source regions corresponding to HMXBs, LMXBs, thermal SNRs, and absorbed sources follow the classification scheme of \citet{2003ApJ...595..719P}. Only observations with constrained measurements in all three energy bands are shown. Error bars represent the propagated uncertainties on the X-ray colors.}
\label{fig:colorcolor}
\end{figure}

\section{X-ray energy spectra of X32 and JWST/NIRCam three-color image of NGC~1300}

\begin{figure}
\resizebox{\hsize}{!}{\includegraphics{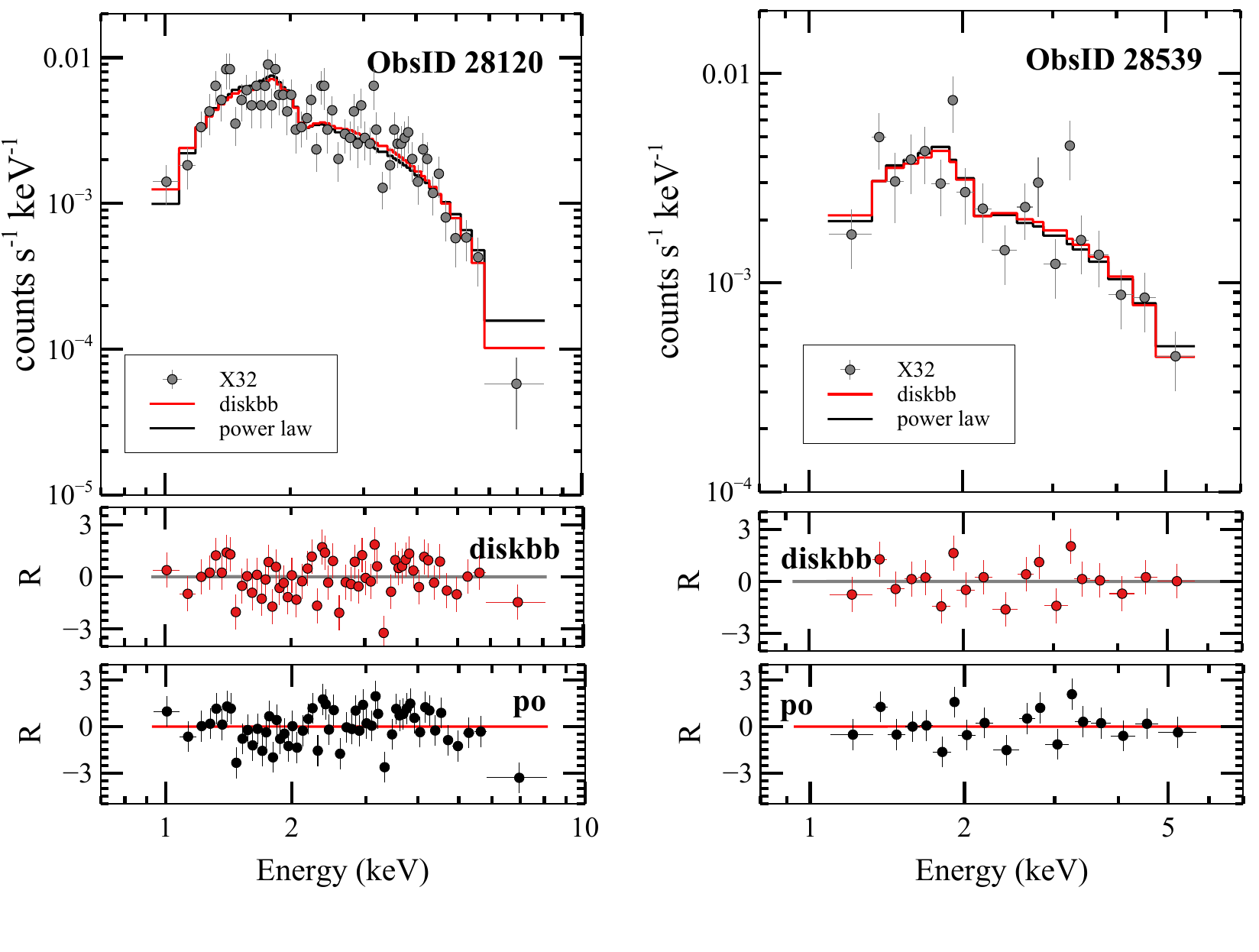}}
\caption{X-ray spectra of X32 obtained from \textit{Chandra} ObsIDs~28120 (left) and 28539 (right). The spectra were fitted with absorbed multicolor disk-blackbody (\textit{TBabs*diskbb}; red) and absorbed power-law (\textit{TBabs*powerlaw}; black) models. The lower panels show the residuals, in units of sigma, for the \textit{diskbb} and power-law fits, respectively.}
\label{fig:modelsX32}
\end{figure}

\begin{figure}
\resizebox{\hsize}{!}{\includegraphics{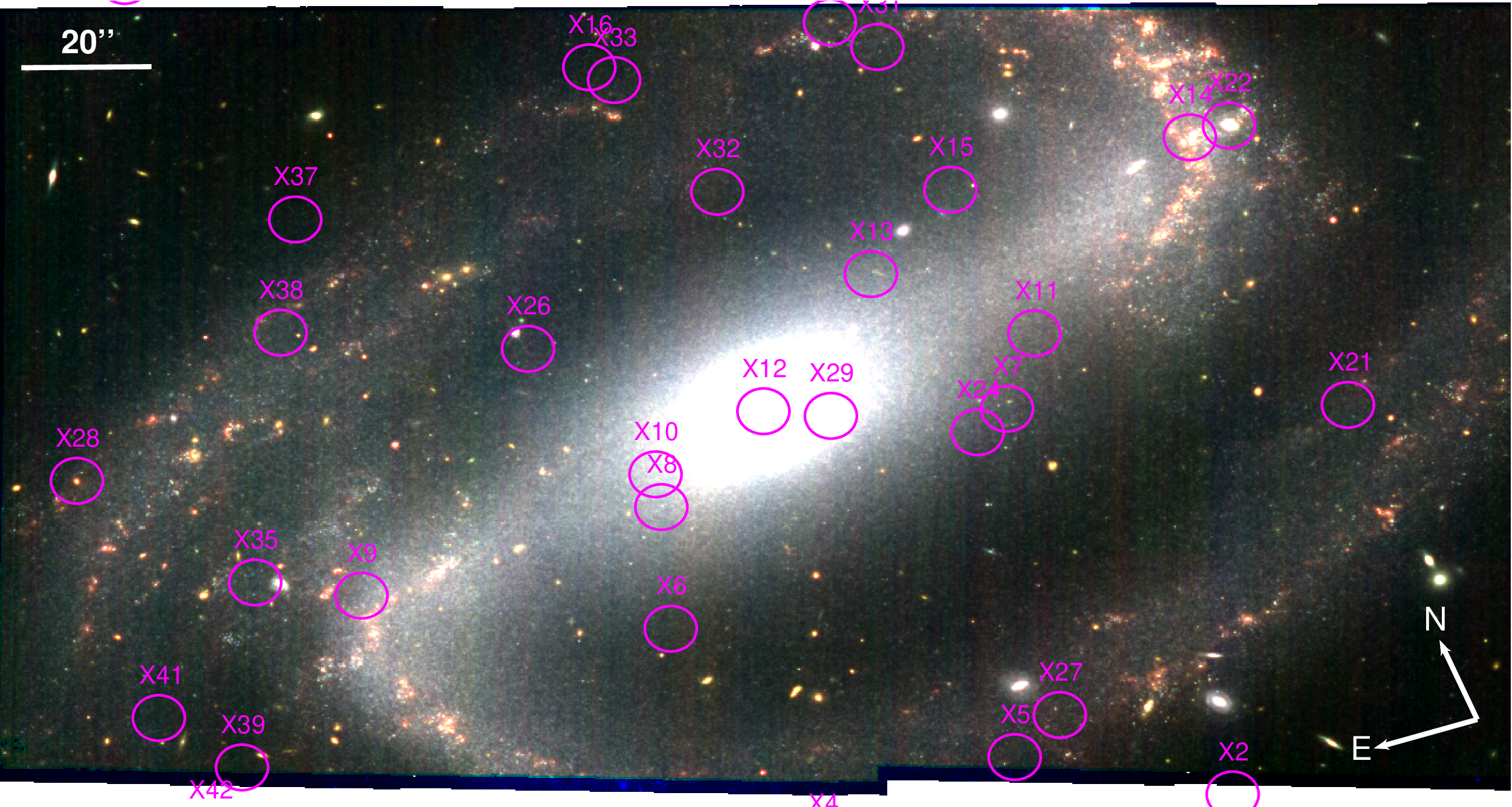}}
\caption{JWST/NIRCam three-color image of NGC~1300 with the positions of
the X-ray sources detected with \textit{Chandra} marked by magenta
circles and labeled X1--X42. The X-ray source population is distributed
across the optical disk, with a substantial fraction of the sources
projected along or close to the prominent spiral-arm structures.
The orientation and angular scale are indicated in the image.}
\label{fig:jwst_rgb}
\end{figure}

\end{document}